\documentclass[fleqn,usenatbib]{mnras}

\usepackage[T1]{fontenc}
\usepackage{ae,aecompl}
\usepackage{multirow}

\usepackage{graphicx}	
\usepackage{amsmath}	
\usepackage{amssymb}	
\usepackage{hyperref}
\usepackage{footmisc} 
\usepackage{newtxtext,newtxmath}

\newcommand{\R}{{\em ROSAT}}

\newcommand{\xmm}{{\em XMM--Newton}}

\newcommand{\RXTE}{{\em RXTE}}
\newcommand{\swift}{{\em Swift}}

\newcommand{\nus}{{\em NuSTAR}}
\newcommand{\nicer}{{\em NICER}}

\def\nh {$N_{\rm H}$}

\def\rchisq {$\chi_{\nu} ^{2}$}
\def\lum {erg\,s$^{-1}$}
\def\flux {erg\,s$^{-1}$cm$^{-2}$}
\def\ss {s\,s$^{-1}$}

\def\cmcm {cm$^{-2}$}

\def\ktcold{$kT_{\rm cold}$}
\def\rcold{$R_{\rm cold}$}

\def\ktmed{$kT_{\rm warm}$}
\def\rmed{$R_{\rm warm}$}

\def\kthot{$kT_{\rm hot}$}
\def\rhot{$R_{\rm hot}$}

\def\srclong{XTE\,J1810--197}
\def\src{XTE\,J1810}

\title[XTE\,J1810--197: the 2018 outburst]{The X-ray evolution and geometry of the 2018 outburst of XTE\, J1810--197}

\author[A. Borghese et al.]{
A. Borghese,$^{1,2}\thanks{E-mail: borghese@ice.csic.es}$
N. Rea,$^{1,2}$ R. Turolla,$^{3,4}$ M. Rigoselli,$^{5}$ J.A.J. Alford,$^{6}$
\newauthor
E.V. Gotthelf,$^{6}$ M. Burgay,$^{7}$ A. Possenti,$^{7,8}$ S. Zane,$^{4}$ F. Coti Zelati,$^{1,2}$ R. Perna,$^{9,10}$  
\newauthor
P. Esposito,$^{11,5}$ S. Mereghetti,$^{5}$ D. Vigan\'o,$^{1,2,12}$ A. Tiengo,$^{11,5,13}$ D. G\"otz,$^{14}$ A. Ibrahim,$^{1,2}$ 
\newauthor
G.L. Israel,$^{15}$ J. Pons,$^{16}$ R. Sathyaprakash$^{1,2}$ 
\\
$^{1}$Institute of Space Sciences (ICE, CSIC), Campus UAB, Carrer de Can Magrans s/n, 08193, Barcelona, Spain\\
$^{2}$Institut d'Estudis Espacials de Catalunya (IEEC), Carrer Gran Capit\`a 2--4, 08034 Barcelona, Spain\\
$^{3}$Dipartimento di Fisica e Astronomia `Galileo Galilei', Universit\`a di Padova, via F. Marzolo 8, 35131 Padova, Italy\\
$^{4}$Mullard Space Science Laboratory, University College London, Holmbury St. Mary, Dorking, Surrey RH5 6NT, UK\\
$^{5}$INAF--Istituto di Astrofisica Spaziale e Fisica Cosmica di Milano, via A.\,Corti 12, I-20133 Milano, Italy \\
$^{6}$Columbia Astrophysics Laboratory, Columbia University, 550 West 120th Street, New York, NY 10027, USA\\
$^{7}$INAF--Osservatorio Astronomico di Cagliari, Via della Scienza 5, 09047 Selargius, Italy\\ 
$^{8}$Department of Physics, Universit\`a di Cagliari, S.P. Monserrato-Sestu km 0,700, 09042 Monserrato, Italy \\
$^{9}$ Department of Physics and Astronomy, Stony Brook University, Stony Brook, NY 11794-3800, USA \\
$^{10}$Center for Computational Astrophysics, Flatiron Institute, 162 5th Avenue, New York, NY 10010, USA \\
$^{11}$Scuola Universitaria Superiore IUSS Pavia, Palazzo del Broletto, piazza della Vittoria 15, 27100 Pavia, Italy \\
$^{12}$Institute of Applied Computing \& Community Code (IAC3), University of the Balearic Islands, Palma de Mallorca, E-07122, Spain \\
$^{13}$Istituto Nazionale di Fisica Nucleare (INFN), Sezione di Pavia, via A.\,Bassi 6, 27100 Pavia, Italy \\
$^{14}$AIM--CEA/DRF/Irfu/Service d'Astrophysique, Orme des Merisiers, F--91191 Gif-sur-Yvette, France\\
$^{15}$INAF--Osservatorio Astronomico di Roma, via Frascati 33, 00078 Monteporzio Catone, Italy \\
$^{16}$Departament de F\'isica Aplicada, Universitat d'Alacant, Ap. Correus 99, E-03080 Alacant, Spain
}

\date{Accepted XXX. Received YYY; in original form ZZZ}

\pubyear{2020}

\begin{document}
\label{firstpage}
\pagerange{\pageref{firstpage}--\pageref{lastpage}}
\maketitle

\begin{abstract}

After 15 years, in late 2018, the magnetar \srclong\ underwent a second recorded X-ray outburst event and reactivated as a radio pulsar. We initiated an X-ray monitoring campaign to follow the timing and spectral evolution of the magnetar as its flux decays using \swift, \xmm, \nus, and \nicer\ observations. During the year-long campaign, the magnetar reproduced similar behaviour to that found for the first outburst, with a factor of two change in its spin-down rate from $\sim7.2\times10^{-12}$\,\ss\ to $\sim1.5\times10^{-11}$\,\ss\, after two months. Unique to this outburst, we confirm the peculiar energy-dependent phase shift of the pulse profile. Following the initial outburst, the spectrum of \srclong\ is well-modelled by multiple blackbody components corresponding to a pair of non-concentric, hot thermal caps surrounded by a cooler one, superposed to the colder star surface. We model the energy-dependent pulse profile to constrain the viewing and surface emission geometry and find that the overall geometry of \srclong\ has likely evolved relative to that found for the 2003 event.
\end{abstract}




\begin{table*}
\caption{Log of the X-ray observations of \srclong\ between 2018 December 13 and 2019 October 25.}
\label{tab:log}
\resizebox{1.5\columnwidth}{!}{
\begin{tabular}{@{}lccccc}
\hline
Obs. ID         &  Instrument$^*$  & Start time (TT) & Mid Point & Exposure & Source net count rate$^{**}$   \\
                &                  &  (YYYY-MM-DD hh:mm:ss) & (MJD) & (ks)    & (counts s$^{-1}$) \\
\hline  
90410368002 & \nus/FPMA & 2018-12-13 03:11:09 & 58465.587 & 39.2 & 1.44(1) \\
90410368002 & \nus/FPMB & 2018-12-13 03:11:09 & 58465.587 & 39.1 & 1.44(1) \\
80202013002 & \nus/FPMA & 2019-01-02 21:06:09 & 58486.346 & 46.0 & 1.13(1) \\
80202013002 & \nus/FPMB & 2019-01-02 21:06:09 & 58486.346 & 45.4 & 1.06(1) \\
80202013003 & \nus/FPMA & 2019-02-07 15:31:09 & 58522.134 & 40.1 & 1.17(1) \\
80202013003 & \nus/FMPB & 2019-02-07 15:31:09 & 58522.134 & 39.7 & 1.15(1) \\
00031335006 & \swift/XRT (WT) & 2019-02-08 19:18:20 & 58522.843 & 1.9 & 3.86(4) \\
00031335007 & \swift/XRT (WT) & 2019-02-15 21:51:38 & 58530.050 & 4.6 & 2.95(3) \\
1020420142 & \nicer & 2019-02-21 02:12:00 & 58535.258 & 1.8 & 46.9(2) \\
00031335008 & \swift/XRT (WT) & 2019-02-22 06:42:33 & 58536.321 & 2.0 & 3.09(4) \\
00031335009 & \swift/XRT (WT) & 2019-03-02 17:24:50 & 58544.764 & 1.5 & 3.07(5) \\
80202013005 & \nus/FPMA & 2019-03-03 22:46:09 & 58546.405 & 38.9 & 0.94(1)  \\
80202013005 & \nus/FPMB & 2019-03-03 22:46:09 & 58546.405 & 38.8 & 0.91(1) \\
2618010101$^a$ & \nicer & 2019-03-03 23:52:25 & 58546.357 & 7.3 & 42.7(1) \\
0784303101 & {\em XMM}/EPIC-pn (TM) & 2019-03-04 02:36:19 & 58546.183 & 12.9 & 29.19(5) \\
2020420103$^a$ & \nicer & 2019-03-04 18:19:12 & 58546.8707 & 4.3 & 43.3(1) \\
00031335010$^\dag$ & \swift/XRT (WT) & 2019-03-08 05:56:17 & 58550.279 & 0.7 & 2.98(7) \\
00031335011$^\dag$ & \swift/XRT (WT) & 2019-03-13 14:43:05 & 58555.652 & 1.5 & 3.14(5) \\
00031335012 & \swift/XRT (WT) & 2019-03-15 22:27:48 & 58558.441 & 2.1 & 3.11(4) \\
00031335013 & \swift/XRT (WT) & 2019-03-22 13:41:45 & 58564.614 & 1.7 & 2.93(4) \\
2020420104 & \nicer & 2019-03-29 19:25:32 & 58571.856 & 3.9 & 39.6(1) \\
00031335014 & \swift/XRT (WT) & 2019-03-29 21:03:31 & 58572.081 & 2.1 & 2.89(4) \\
2020420105$^b$ & \nicer & 2019-03-31 17:47:14 & 58573.853 & 7.5 & 37.5(1) \\ 
2020420106$^b$ & \nicer & 2019-03-31 23:58:33 & 58574.014 & 2.0 & 37.6(1) \\
00031335015$^\ddag$ & \swift/XRT (WT) & 2019-04-05 04:25:28 & 58578.260 & 1.0 & 2.14(5) \\
2020420107 & \nicer & 2019-04-07 13:42:40 & 58580.618 & 3.4 & 36.5(1) \\
2618010201$^c$ & \nicer & 2019-04-12 16:12:40 & 58585.812 & 3.7 & 36.4(1) \\
00031335016$^\ddag$ & \swift/XRT (WT) & 2019-04-12 21:40:54 & 58585.932 & 0.6 & 2.43(7) \\
2618010202$^c$ & \nicer & 2019-04-12 23:56:40 & 58586.167 & 3.9 & 36.4(1) \\
00031335017 & \swift/XRT (WT) & 2019-04-17 06:41:10 & 58590.584 & 1.5 & 2.69(4) \\
00031335018 & \swift/XRT (WT) & 2019-04-19 01:42:46 & 58592.136 & 1.9 & 2.77(4) \\
00031335021 & \swift/XRT (WT) & 2019-05-03 13:10:44 & 58606.585 & 1.4 & 2.51(4) \\
00031335022 & \swift/XRT (WT) & 2019-05-10 18:47:11 & 58614.313 & 1.8 & 0.48(2) \\
00031335023 & \swift/XRT (WT) & 2019-05-17 15:08:39 & 58620.733 & 1.6 & 1.55(3) \\
2618010301$^d$ & \nicer & 2019-05-23 21:41:40 & 58626.947 & 2.8 & 31.9(1) \\
2618010302$^d$ & \nicer & 2019-05-24 00:47:40 & 58627.495 & 9.9 & 31.9(1) \\
00031335024 & \swift/XRT (WT) & 2019-05-24 18:49:41 & 58627.819 & 1.8 & 2.25(4) \\
2618010303 & \nicer & 2019-05-25 00:12:00 & 58628.337 & 4.1 & 31.7(1) \\
00031335025$^e$ & \swift/XRT (WT) & 2019-05-30 15:02:56 & 58633.634 & 0.4 & 2.25(7) \\
00031335026$^e$ & \swift/XRT (WT) & 2019-06-01 19:57:34 & 58635.904 & 1.6 & 1.96(4) \\
00031335027 & \swift/XRT (WT) & 2019-06-09 06:18:32 & 58643.302 & 1.9 & 2.16(3) \\
00031335028$^f$ & \swift/XRT (WT) & 2019-06-25 17:23:02 & 58659.729 & 0.9 & 1.96(4) \\
00031335029$^f$ & \swift/XRT (WT) & 2019-06-26 12:31:41 & 58660.525 & 0.6 & 2.12(6) \\
00031335031$^{\dag\dag}$ & \swift/XRT (WT) & 2019-06-30 23:43:59 & 58664.992 & 0.5 & 1.94(6) \\
00031335032$^{\dag\dag}$ & \swift/XRT (WT) & 2019-07-03 12:11:51 & 58667.513 & 0.8 & 1.88(5) \\
2618010401 & \nicer & 2019-07-07 01:25:30 & 58671.493 & 7.9 & 27.4(1) \\
00031335034 & \swift/XRT (WT) & 2019-07-12 04:46:47 & 58676.272 & 1.1 & 1.67(4) \\
00031335036 & \swift/XRT (WT) & 2019-07-25 18:14:18 & 58689.828 & 1.5 & 1.14(2) \\
00031335037 & \swift/XRT (WT) & 2019-08-01 00:03:15 & 58696.002 & 0.9 & 0.97(3) \\
2618010501$^g$ & \nicer & 2019-08-05 20:57:40 & 58700.908 & 1.0 & 24.4(2) \\ 
2618010502$^g$ & \nicer & 2019-08-06 00:03:20 & 58701.489 & 2.9 & 23.7(1) \\
2618010503$^g$ & \nicer & 2019-08-07 00:31:47 & 58702.480 & 3.5 & 23.5(1) \\
2618010504 & \nicer & 2019-08-07 23:42:47 & 58703.482 & 6.5 & 23.6(1) \\
2618010505 & \nicer & 2019-08-14 01:09:00 & 58709.215 & 1.1 & 23.8(1) \\
00031335038 & \swift/XRT (WT) & 2019-08-13 15:54:19 & 58708.700 & 1.6 & 1.74(3) \\
00031335043 & \swift/XRT (PC) & 2019-09-17 01:28:53 & 58743.530 & 1.2 & 0.57(2) \\
2618010601 & \nicer & 2019-09-21 10:18:25 & 58747.695 & 6.0 & 21.3(1) \\
2618010602 & \nicer & 2019-09-22 00:14:25 & 58748.308 & 6.9 & 21.3(1) \\
30501023002 & \nus/FPMA & 2019-09-22 00:16:09 & 58749.212 & 95.6 & 0.45(2) \\
30501023002 & \nus/FPMB & 2019-09-22 00:16:09 & 58749.212 & 96.1 & 0.44(2) \\ 
0784303201 & {\em XMM}/EPIC-pn (SW) & 2019-09-22 14:52:41 & 58748.947 & 56.5 & 9.49(1) \\
00031335044 & \swift/XRT (PC) & 2019-09-25 18:14:20 & 58751.864 & 1.6 & 0.57(2) \\
30501023004 & \nus/FPMA & 2019-09-26 21:51:09 & 58753.481 & 43.2 & 0.42(3) \\
30501023004 & \nus/FPMB & 2019-09-26 21:51:09 & 58753.481 & 43.5 & 0.41(3) \\
00031335045 & \swift/XRT (PC) & 2019-10-06 21:59:58 & 58762.954 & 1.6 &  0.40(2) \\ 
00031335046 & \swift/XRT (PC) & 2019-10-15 00:35:43 & 58771.098 & 2.2 & 0.49(1) \\
00031335047 & \swift/XRT (PC) & 2019-10-25 04:21:03 & 58781.215 & 1.7 & 0.47(2) \\
\hline
\end{tabular}
}
\begin{list}{}{}
\item[$^*$] The instrumental setup is indicated in brackets: WT = windowed timing, TM = timing mode, PC = photon counting and SW = small window.
\item[$^{**}$] The source net count rate in the 0.3--10\,keV range for \xmm\ and \swift, in the 0.6--7\,keV interval for \nicer, and in the 3--15\,keV range for \nus. 
\item[$^{a,b,c,d,e,f,g}$] These observations were merged in the spectral analysis.
\item[$^{a,\dag, b,\ddag,c,d,e,f,\dag\dag, g}$] These observations were merged in the timing analysis.
\end{list}
\end{table*}


\section{Introduction}

Magnetars are isolated pulsars whose emission is thought to be powered by the decay and instabilities of their extreme magnetic fields, typically $B$ $\sim$ 10$^{14}$--10$^{15}$~G \citep[see e.g.,][for reviews]{2017ARA&A..55..261K,2021ASSL..461...97E}. With spin periods in the 0.3--12\,s range and relatively large spin-down rates, these objects have a persistent X-ray luminosity of $L_X$ $\sim$ 10$^{31}$--10$^{36}$\,\lum, generally larger than their rotational energy loss rate. The main feature of these isolated neutron stars is the unpredictable and variable bursting activity in the X-/gamma-ray bands on different time scales. These flaring events often indicate that the source has entered an active phase, commonly referred to as outburst \citep[see the Magnetar Outburst Online Catalog,][]{2018MNRAS.474..961C}. During an outburst, the persistent X-ray flux increases by up to three orders of magnitude above the quiescent level. Then, it usually relaxes back to the pre-outburst level on time scales spanning from weeks to months/years. Up to now, about 30 sources are listed as magnetars, although magnetar-like activity has been recorded from other classes of isolated neutron stars too, such as high-$B$ radio pulsars \citep[e.g.,][]{2016ApJ...829L..21A} and the source at the centre of the supernova remnant RCW\,103 \citep{2016MNRAS.463.2394D, 2016ApJ...828L..13R}.\\ 

\srclong\ (\src) unveiled its magnetar nature in 2003 when the {\em Rossi X-ray Timing Explorer} (\RXTE) detected an increase in its X-ray flux by a factor of $\sim$ 100 with respect to the quiescent level measured by \R\ in 1993, $\sim$5$\times10^{-13}$~\flux\ \citep[0.5--10~keV;][]{2004ApJ...609L..21I,2004ApJ...605..368G}. The outburst provided the opportunity to detect the spin signal, at a period of $\sim$ 5.54~s. 
Archival VLA 1.4\,GHz survey data from 2004 January revealed post-outburst radio emission from the magnetar \citep{2005ApJ...632L..29H} that lead to the discovery in 2006 of radio pulsations at the X-ray spin period \citep{2006Natur.442..892C}. \src\ was the first magnetar to show pulsed emission in the radio band. The X-ray flux reached the pre-outburst level in early 2007 \citep{2009A&A...498..195B,2016ApJ...818..122A}, although the source remained radio loud until late 2008 \citep{2016ApJ...820..110C}.  
The initial phases of the outburst were missed, therefore it was not possible to fully characterize the magnetar properties at the outburst peak. \xmm\ started to monitor \src\ approximately one year after the outburst onset. Spectral and temporal studies during the early decay revealed a variable spin-down rate and a two temperature thermal spectrum, suggesting a localized surface hot spot with temperature of \kthot$\sim$0.7\,keV surrounded by a cooler corona with \ktmed$\sim$0.3\,keV \citep[see, e.g.,][and references therein]{2005ApJ...618..874H}.
These two regions were superimposed on the quiescent component ($kT_{\rm cold}\sim$0.15\,keV) which was identified with the thermal emission from the whole neutron star surface \citep{2008ApJ...681..522P,2009A&A...498..195B,2010ApJ...722..788A,2016ApJ...818..122A}. An extensive radio and X-ray monitoring allowed to study the timing properties of \src\ over the outburst evolution until the quiescent state. The period derivative was highly variable during the outburst decay, while it remained steady in quiescence, $\sim$3$\times10^{-12}$\,\ss\ \citep{2016ApJ...820..110C, 2016MNRAS.458.2088P, 2019MNRAS.483.3832P}.

After 11 yr of low activity in both the radio and X-ray band, \src\ showed intense radio emission and an X-ray enhancement on 2018 December 8 \citep{2018ATel12284....1L,2018ATel12291....1M}.
A more detailed analysis constrained the onset of the new outburst between 2018 October 26 and 2018 December 8 at radio wavelengths and between 2018 November 20--26 for the X-ray energy band \citep{2019ApJ...874L..25G}. Given the proximity of the source to the Sun at that epoch, observations in the soft X-ray band were not feasible till 2019 February. However, the {\em Nuclear Spectroscopic Telescope Array} (\nus) observed the source on 2018 December 13 and detected X-ray emission up to $\sim$ 30\,keV \citep{2019ApJ...874L..25G}. The 3--10\,keV spectrum was well described by a blackbody ($kT \sim$0.74\,keV) plus power-law (photon index $\Gamma \sim$4.3) model. To account for the non-thermal emission above 10\,keV, an additional power law was required ($\Gamma_{\rm h} \sim -$0.31). The 2--10\,keV absorbed flux of 2$\times10^{-10}$\,\flux\ was a factor of $\sim$2 higher than the maximum flux extrapolated for the 2003 outburst.    

In this work, we present a monitoring campaign of the 2018 outburst decay of \src\ carried out with \nus, \xmm, the {\em Neil Gehrels Swift Observatory} (\swift) and {\em Neutron Star Interior Composition Explorer} (\nicer) covering $\sim$11 months since 2018 December. We describe the details of the X-ray data analysis in Section \ref{sect:data} and present results in Section \ref{sect:results}. In Section \ref{sect:radio}, we report on a radio observation performed with the Sardinia Radio Telescope, simultaneously with one of the \nus\ pointings. Finally, implications are discussed in Section \ref{sect:disc}. Conclusions follow is Section \ref{sect:conc}.\\  

\section{X-ray observations and data reduction}
\label{sect:data}

Our campaign followed \src\ from 2018 December 13 till 2019 October 25. A log of the observations analysed in this paper is reported in Table \ref{tab:log} (for completeness, we included in our analysis the \nus\ pointing already reported by \citealt{2019ApJ...874L..25G}). Data reduction was performed using tools incorporated in {\sc heasoft} (version 6.27) and the Science Analysis Software ({\sc sas}, version 18).

Through this work, we adopt a distance of 2.5\,kpc derived using radio parallax \citep{2020MNRAS.498.3736D}. Photon arrival times for all satellites were barycentered using the radio position of \src, $\rm RA = 18^h 09^m 51\fs087$, $\rm Decl = -19^\circ 43' 51\farcs93$ \citep[J2000;][]{2007ApJ...662.1198H}, and the JPL planetary ephemeris DE\,200. In the following, the uncertainties are quoted at 1$\sigma$ confidence level (c.l.).

\begin{figure}
    \centering
    \includegraphics[width=1.\columnwidth]{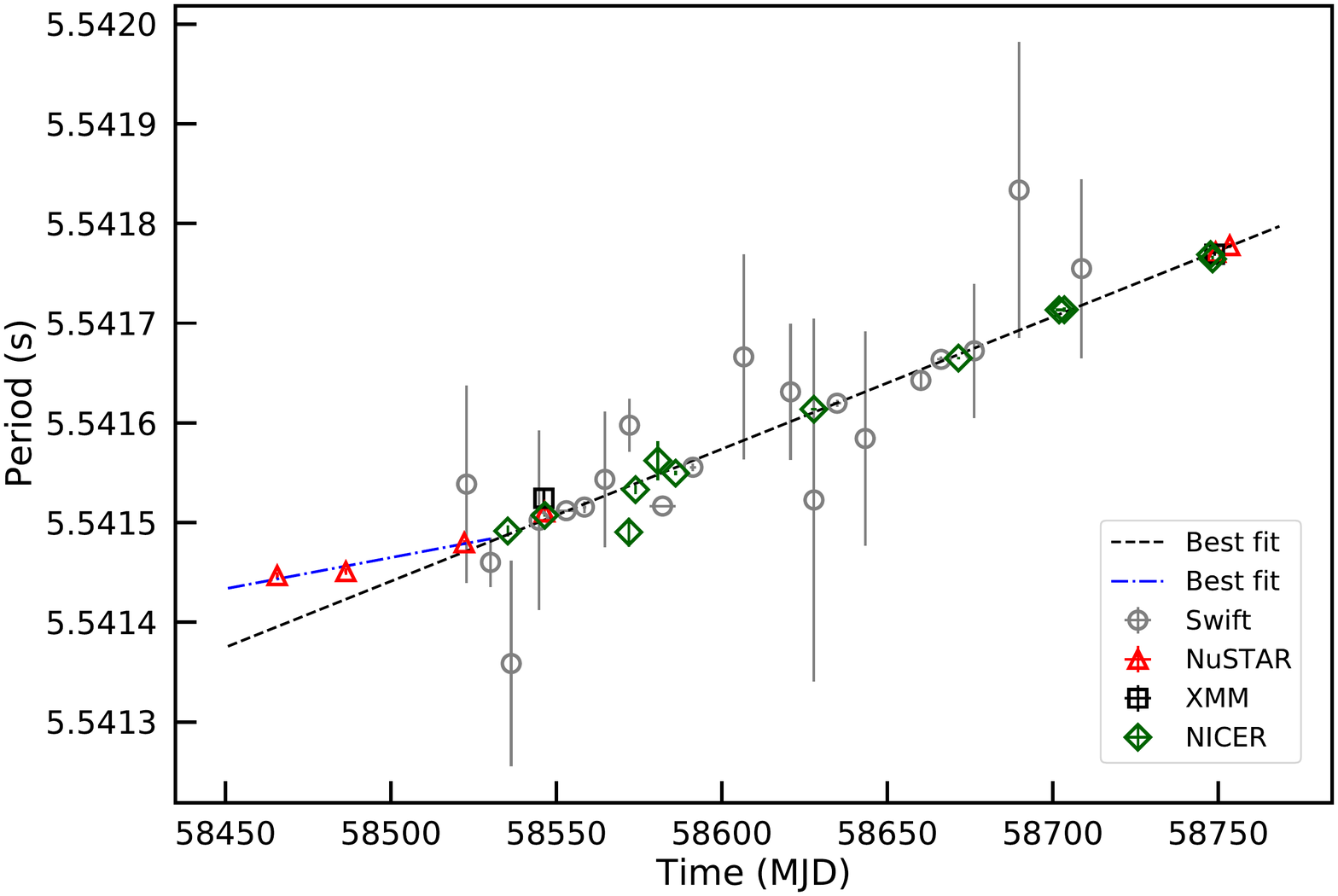}
    \caption{Values for the spin period along the 2018 outburst of \srclong\ as measured in the single X-ray observations. Grey circles, red triangles, black squares and green diamonds are the measurements for \swift, \nus, \xmm\ and \nicer\ pointings, respectively. The black dashed and blue dash-dotted lines indicate the best fit for all the data sets and a sub-sample between 2018 December 13 and 2019 February 7.}
    \label{fig:period}
\end{figure}

\subsection{\nus}

\src\ was observed by \nus\ six times, for a total exposure time of $\sim$300\,ks. \nus\ consists of two co-aligned X-ray telescopes, which focus onto two focal plane detectors, referred to as FPMA and FPMB, sensitive to photons in the 3--79\,keV energy band \citep{2013ApJ...770..103H}. The calibrated timing accuracy of \nus\ is $\sim$65\,$\mu$s, after correcting
for drift of the on-board clock \citep{2021ApJ...908..184B}.

For observations IDs 90410368002 and 80202013002, the small angular separation between the source and the Sun did not allow the star tracker co-aligned with the X-ray optics to provide an aspect solution. However, an approximate solution could be reconstructed using a combination of the other three star trackers available. Event files corresponding to each star tracker configuration were generated using the {\sc runsplitsc} option in {\sc nupipeline}. We extracted light curves and background-subtracted spectra, and created redistribution matrices and ancillary files for each event file. Then, we combined them by means of the tool {\sc addascaspec} to generate the averaged files for both FPMs. For the other pointings, we applied standard analysis threads. Source counts were accumulated within a circular region with radius 50 arcsec, while background events were extracted from different regions (e.g., a circle of radius 100 arsec or an annulus with radii of 100 and 150 arcsec). We checked that the different choices for the background regions yielded consistent results in the spectral and timing analysis. 

\subsection{\xmm}
\xmm\ observed \src\ twice, on 2019 March 4 and September 22, with the European Photon Imaging Cameras (EPIC) for a total exposure time of $\sim$70~ks. The EPIC-pn \citep{2001A&A...365L..18S} was set in timing mode (TM; timing resolution of 0.03\,ms) during the first pointing and in small window mode (SW; 5.7\,ms) for the second one. For both observations, the MOS1 camera was operating in full frame mode (timing resolution of 2.6\,s) while the MOS2 in timing mode \citep[1.75\,ms;][]{2001A&A...365L..27T}. These pointings were coordinated with \nus\ in order to probe the magnetar emission over a broader energy range due to \xmm\ high sensitivity to soft X-rays (0.3--10\,keV). In this work, we considered only data acquired with the EPIC-pn camera, because it provides the spectra with the highest counting statistics owing to its larger effective area.

Raw data were processed via the {\sc epproc} task. We cleaned the observations from any particle flares collecting light curves above 10\,keV and employing an intensity threshold. In the first observation, we collected source photons from a strip with width of 11 pixels and background counts far away from the source within a strip with width of 2 pixels (1 pixel = 4.1 arcsec). While in the second pointing, the source events were selected from a 40-arcsec radius circle and the background was extracted from a source-free circular region of radius 60 arcsec. Following standard analysis procedure, the response matrices and ancillary files were generated through the {\sc rmfgen} and {\sc arfgen} tools, respectively. 

\subsection{\swift}

\src\ was monitored by the X-ray Telescope \citep[XRT;][]{2005SSRv..120..165B} on board \swift, for a total exposure of $\sim$50\,ks. The single exposure times ranged from 0.4 to 4.6\,ks, with the XRT operating either in windowed timing (WT;  1.77\,ms) or photon counting modes (PC; 2.51\,s).

We reprocessed the data adopting standard cleaning criteria and created exposure maps with the task {\sc xrtpipeline}. For the spectral analysis, we selected events with grades 0--12 and 0 for PC and WT data, respectively, while we extended the timing analysis to events with grades 0--2 for WT data sets. We accumulated the source counts from a circular region with radius 20 pixels (1 pixel = 2.36 arcsec). To evaluate the background in PC data, we extracted the events within an annulus centred on the source position with radii of 40 and 80 pixels. For the WT data, we adopted a region far from the target and of the same size as that used for the source. In case an observation performed in PC mode was affected by pile-up, we followed the online analysis thread\footnote{\url{https://www.swift.ac.uk/analysis/xrt/pileup.php}} to determine 
the size of the core of the point-spread function to be excluded from our analysis. We generated the spectra with the corresponding ancillary response files through {\sc xselect} and the {\sc xrtmkarf} tool. The response matrices version `20131212v015' and `20130101v014' available in the XRT calibration database were assigned to each spectrum in WT and PC mode, respectively. 

\subsection{\nicer}
\nicer\ was installed on the International Space Station in June 2017 \citep{2012SPIE.8443E..13G}. The payload, the X-ray Timing Instrument (XTI), consists of 56 co-aligned collimators that allow X-rays in the 0.2--12\,keV range onto paired silicon drift detectors. The 52 functioning detectors provide an effective area of $\sim$1900\,cm$^2$ at 1.5\,keV. \nicer\ combines good energy resolution ($\sim$100\,eV at 1\,keV) with an excellent timing resolution ($<$300\,ns), being thus the perfect instrument to perform spectral and timing studies.     

\nicer\ extensively monitored \src\ from the beginning of the mission till 2018 July. When the latest outburst took place, the source was in a Solar constrained period. Observations resumed in 2019 February and covered the whole year. We included only observations with an exposure time longer than 1\,ks in our analysis, for a cumulative exposure of 90.5\,ks. The data were processed via the {\sc nicerdas} pipeline, with the tool \textsc{nicerl2} and using standard filtering criteria. Since \nicer\ does not provide imaging capabilities, the background count rate and spectra are computed from \nicer\ observations of the \RXTE\ blank-field regions using the {\sc nibackgen3C50} tool. The response matrix `20170601v002' and ancillary file `20170601v004' were assigned to the background-subtracted spectra with the {\sc grppha} task.

\section{X-ray analysis and results}
\label{sect:results}

\begin{figure}
    \centering
    \includegraphics[width=1.\columnwidth, trim = 3cm 0 3cm 4cm, clip]{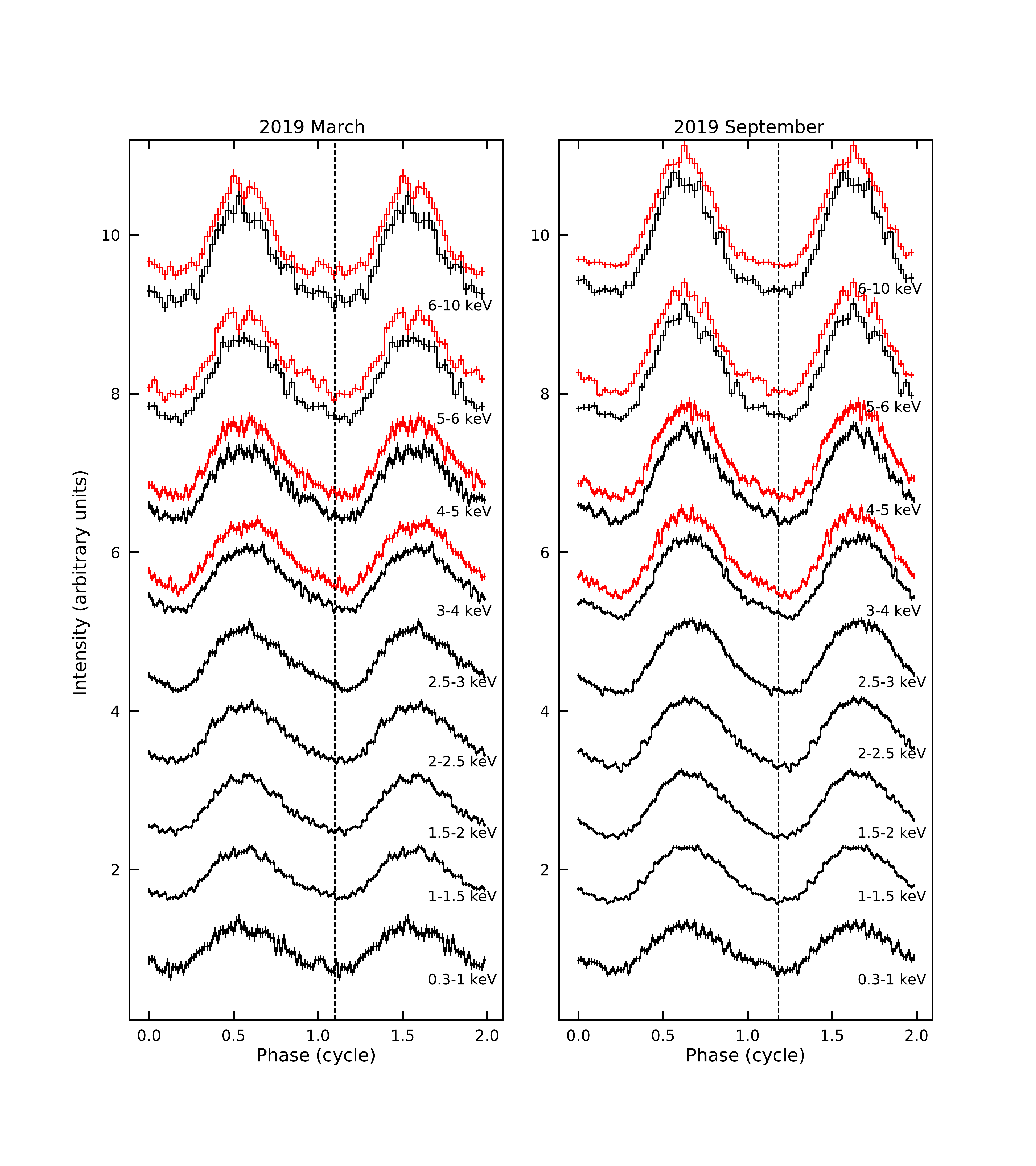}
     \vspace{-28pt}
    \caption{Energy-resolved background-subtracted pulse profiles from \xmm\ (black) and \nus\ (red) data sets for 2019 March (left) and September (right). The dashed lines indicate the minimum of the 0.3--1\,keV pulse profiles in the two cases.}
    \label{fig:pulseperiod}
\end{figure}

\subsection{Timing analysis}
\label{sect:timing}

In order to study the evolution of the spin period during the outburst decay, we selected events in the 0.3--10\,keV energy band for \xmm\ and \swift, 0.6--7\,keV for \nicer\ and 3--10\,keV for \nus. For the latter, we combined the FPMA and FPMB event files for each observation. Observations performed a few days apart were merged to increase the source signal-to-noise ratio (see Table\,\ref{tab:log}). 

We initially extrapolated the timing solution provided by \citet{2019ApJ...877L..30G}, who used \nicer\ data covering the timespan 2019 February 07--15 (MJDs 58521.043--58529.490). The predicted spin period at the epoch of our data sets was employed as trial period in an epoch folding search technique, which gave us a more precise value of the period. Finally, we refined these values by applying a phase-fitting procedure in each observation \citep[for more details see, e.g.,][]{2003ApJ...599..485D}. Within each pointing, we were able to align the pulse phases using only a linear function. The periods derived following these steps are plotted as a function of time in Figure \ref{fig:period}. In order to constrain the long-term average spin-down rate, 
we fit the period evolution with a first order polynomial function obtaining a slope of 1.5(3) $\times$ 10$^{-11}$\,\ss. This value is consistent within 1$\sigma$ with the period derivative derived during the first three years of the 2003 outburst with the same technique \citep{2016MNRAS.458.2088P}, and a factor of $\sim$5 higher than the value measured in quiescence (2.84(2) $\times$ 10$^{-12}$\,\ss; \citealt{2019MNRAS.483.3832P}). 

A phase-coherent timing solution was reported for the initial phase of the outburst in the radio band with a baseline of $\sim$50 days, from 2018 December 8 till 2019 January 24 \citep[see Table 1 in][]{2019MNRAS.488.5251L}. We tried to phase-connect all the observations, but this task was not possible due to the large timing noise. In order to compare the radio measurement, $\dot{P}$=7.91(1) $\times$ 10$^{-12}$\,\ss, with our results, we modeled the period temporal evolution between 2018 December 13 and 2019 February 7 with a linear function, deriving a spin-down rate of (7.2$\pm$1.5) $\times$ 10$^{-12}$\,\ss, value close to the radio estimate. 
By fitting the rest of the measurements, we obtained a spin-down rate of $\sim$1.5 $\times$ 10$^{-11}$\,\ss, consistent with the long-term average spin-down we derived from the entire monitoring campaign and roughly twice as large as the value inferred from the first months of the outburst (see \S\ref{sect:disc}).
 
\begin{figure}
    \centering
    \includegraphics[width=1.\columnwidth, trim = 2.5cm 0 3cm 4cm, clip]{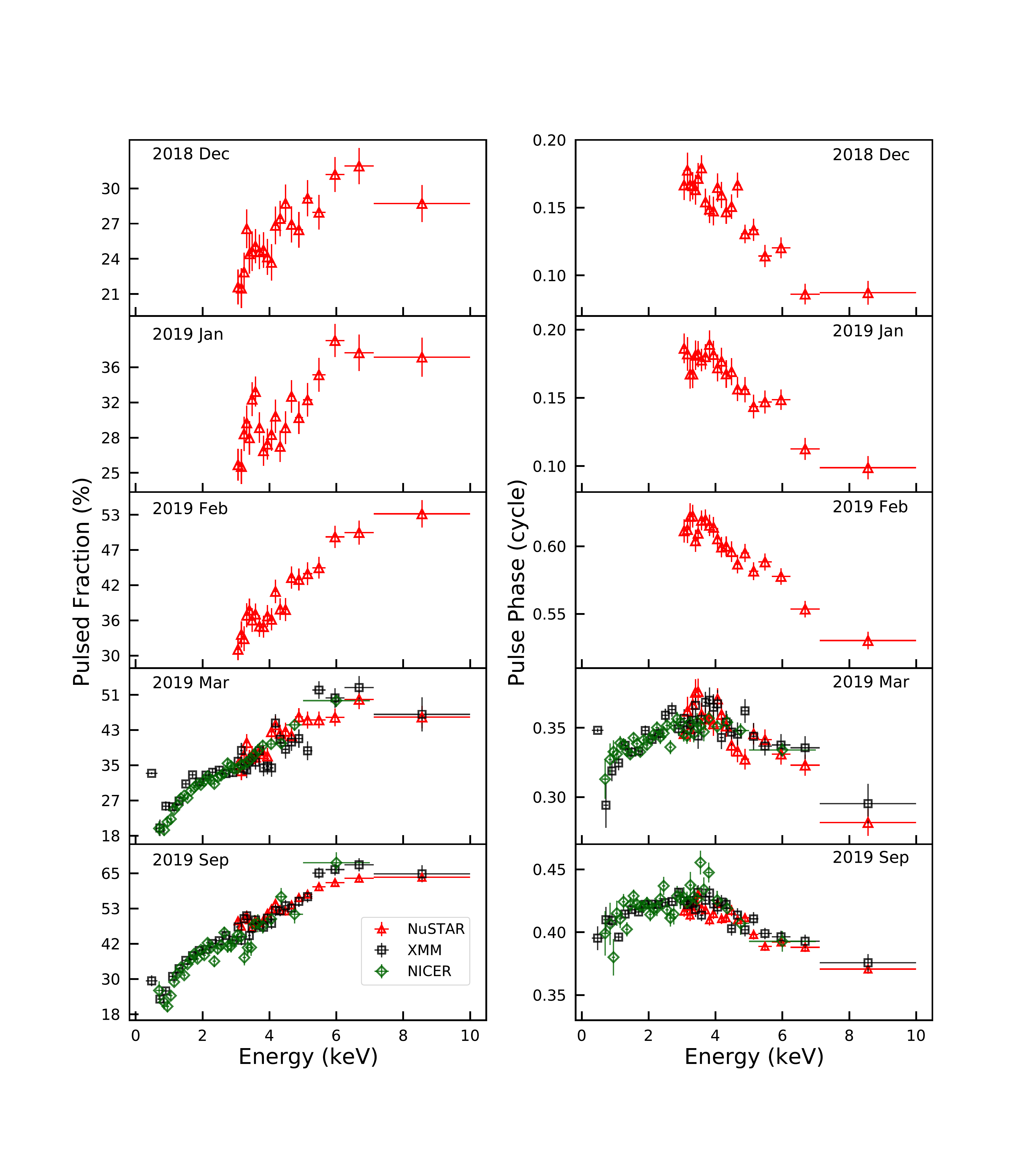}
    \vspace{-30pt}
    \caption{
    Pulsed fraction (left) and pulse phase of the fundamental component (right) as a function of energy for \nus\ observations (red triangles) and the (quasi-)simultaneous \xmm\ (black squares) and \nicer\ (green diamonds) pointings. We note that due to the lack of a phase-connected timing solution, the pulse profiles are not aligned with each others.}
    \label{fig:pfshift}
\end{figure}

By folding each light curve at its measured spin period, we obtained the pulse profile and found that it remained single-peaked in all observations (see Figure\,\ref{fig:pulseperiod}).
We modelled all the pulse profiles with a combination of a constant plus two sinusoidal functions, with periods fixed to those of the fundamental and first harmonic ($F$-test probability $>10^{-3}$ for the inclusion of the second sinusoidal component). We studied the dependence of the pulsed fraction (PF) with the photon energy and its evolution along the outburst decay. The PF was computed by dividing the value of the semi-amplitude for the fundamental sinusoidal component describing the pulse profile by the average count rate. Between 2018 December and 2019 October, the PF increased by a factor of $\sim$2 (from $\sim$27\% to $\sim$54\%) in the 3--10\,keV energy interval. Moreover, the PF showed an increase as a function of energy in every observation (see Figure \ref{fig:pfshift}, left panel) and as a function of time in the same energy band. Figure \ref{fig:pfshift}, right panel, shows the pulse phase of the fundamental component as a function of energy for (quasi-)simultaneously \nus, \xmm\ and \nicer\ observations. We clearly detected the slippage in phase reported in \citet{2019ApJ...874L..25G}, with a magnitude of $\sim$0.1 phase cycles in the first four epochs and $\sim$0.05 in the last one. While in the 0.3--3.5\,keV band, the pulse phase increased with energy by $\sim$0.1 cycles in 2019 March and $\sim$0.04 in 2019 September. 

\begin{table*}
\begin{center}
\caption{Results of the joint fit with a 3BB model of the \xmm\ and \nus\ spectra extracted from the nearly simultaneous observations performed on 2019 March and September (see Section \ref{sect:spectral}).}
\label{tab:spectra_3bb}
\begin{tabular}{|l|ccccccc}
\hline
\hline
Epoch & \ktcold & \rcold & \ktmed & \rmed & \kthot & \rhot & \rchisq/dof \\
      &  (keV)  &  (km)  & (keV)  & (km)  &  (keV) &  (km) &  \\
\hline
2019 Mar & 0.26$\pm$0.01 & 5.4$\pm$1.1 & 0.702$\pm$0.004 & 1.9$\pm$0.3 & 1.83$\pm$0.07 & 0.07$\pm$0.01 & \multirow{2}{*}{1.07/613} \\
2019 Sep & 0.22$\pm$0.01 & 5.6$\pm$1.2 & 0.669$\pm$0.004 & 1.4$\pm$0.2 & 1.48$\pm$0.08 & 0.08$\pm$0.01 & \\
\hline
\hline
\end{tabular}
\end{center}
{\bf Notes.} The derived absorption column density was \nh\ = (8.7$\pm$0.3)$\times10^{21}$\,\,cm$^{-2}$.
\end{table*}

\begin{table*}
\begin{center}
\caption{Results of the joint fit with a 4BB model of the \xmm\ and \nus\ spectra extracted from the nearly simultaneous observations performed on 2019 March and September (see Section \ref{sect:spectral}).}
\label{tab:spectra_4bb}
\begin{tabular}{|l|ccccccc}
\hline
\hline
Epoch & \ktcold & \rcold & \ktmed & \rmed & \kthot & \rhot & \rchisq/dof \\
      &  (keV)  &  (km)  & (keV)  & (km)  & (keV)  &  (km) &  \\
\hline
2019 Mar & 0.26$\pm$0.01 & 5.9$\pm$1.1 & 0.699$\pm$0.004 & 1.9$\pm$0.3 & 1.81$\pm$0.08 & 0.07$\pm$0.01 & \multirow{2}{*}{1.06/613} \\
2019 Sep & 0.25$\pm$0.01 & 4.5$\pm$0.9 & 0.669$\pm$0.004 & 1.4$\pm$0.2 & 1.49$\pm$0.05 & 0.07$\pm$0.01 & \\
\hline
\hline
\end{tabular}
\end{center}
{\bf Notes.} The temperature and the radius of one BB were fixed to the values of the quiescent component, $R_{\rm NS}$ = 12.8\,km and $kT_{\rm NS}$ =  0.144\,keV \citep{2009A&A...498..195B}. The derived absorption column density was \nh\ = (9.7$\pm$0.2)$\times$10$^{21}$\,cm$^{-2}$. 
\end{table*}

\begin{table*}
\begin{center}
\caption{Results of the fit with a 4BB model of the \nus\ spectra (see Section \ref{sect:spectral}).}
\label{tab:spectra_4bb_nustar}
\begin{tabular}{|l|cccccc}
\hline
\hline
Epoch &  \ktmed & \rmed & \kthot & \rhot & Flux & \rchisq/dof \\
      &  (keV)  & (km)  & (keV)  & (km)  & (10$^{-11}$\,\flux)     &    \\
\hline
2019 Jan & 0.66$\pm$0.01 & 2.6$\pm$0.4 & 1.49$\pm$0.05 & 0.16$\pm$0.03 & 14.5$\pm$0.5 & 1.1/146 \\
2019 Feb & 0.68$\pm$0.01 & 2.4$\pm$0.4 & 1.50$\pm$0.06 & 0.14$\pm$0.03 & 13.5$\pm$0.4 & 0.99/139 \\ 
2019 Mar & 0.63$\pm$0.01 & 2.5$\pm$0.4 & 1.38$\pm$0.06 & 0.17$\pm$0.04 & 11.5$\pm$0.2 & 0.99/131 \\
2019 Sep & 0.64$\pm$0.01 & 1.6$\pm$0.3 & 1.40$\pm$0.05 & 0.09$\pm$0.02 & 4.9$\pm$0.1 & 1.1/118 \\
\hline
\hline
\end{tabular}
\end{center}
{\bf Notes.} Two BB component parameters were frozen to $R_{\rm NS}$=12.8\,km, $kT_{\rm NS}$=0.144\,keV, \ktcold=0.26\,keV and \rcold=5.9\,km. The absorption column density was fixed at \nh=9.7$\times$10$^{21}$\,cm$^{-2}$. The flux is the observed flux estimated in the 0.3--10\,keV energy range. 
\end{table*}

\subsection{Phase-averaged spectral analysis}
\label{sect:spectral}

We binned the \nus\ and \xmm\ energy spectra to guarantee at least 100 background-subtracted counts per bin. The \swift/XRT background-subtracted spectra were grouped according to a minimum number of counts variable from observation to observation, varying between 20 and 50 counts per spectral bin, while for \nicer\ spectra we adopted 500 background-subtracted counts per bin. 

The spectral analysis was performed using {\sc Xspec} \citep[version 12.10.1f;][]{1996ASPC..101...17A}. Once the best fit for the adopted model was found, the observed and unabsorbed fluxes were estimated with the convolution model {\sc cflux}. Photoelectric absorption by the interstellar medium was included using the {\sc Tbabs} model with photoionization cross-section from \citet{1996ApJ...465..487V} and chemical abundances from \citet*{2000ApJ...542..914W}.

\begin{figure}
   \centering
   \includegraphics[width=1.\columnwidth]{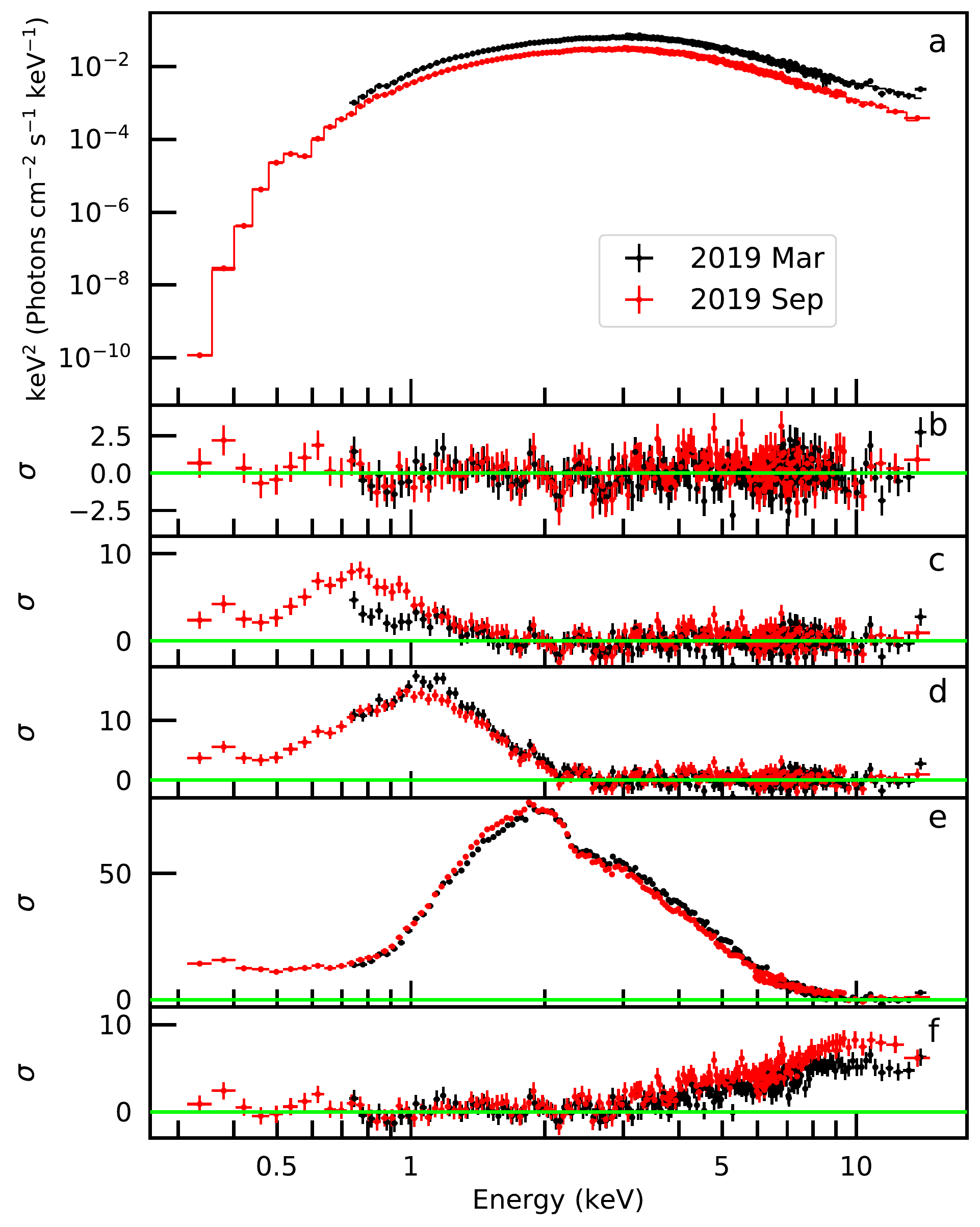}
   \includegraphics[width=1.\columnwidth]{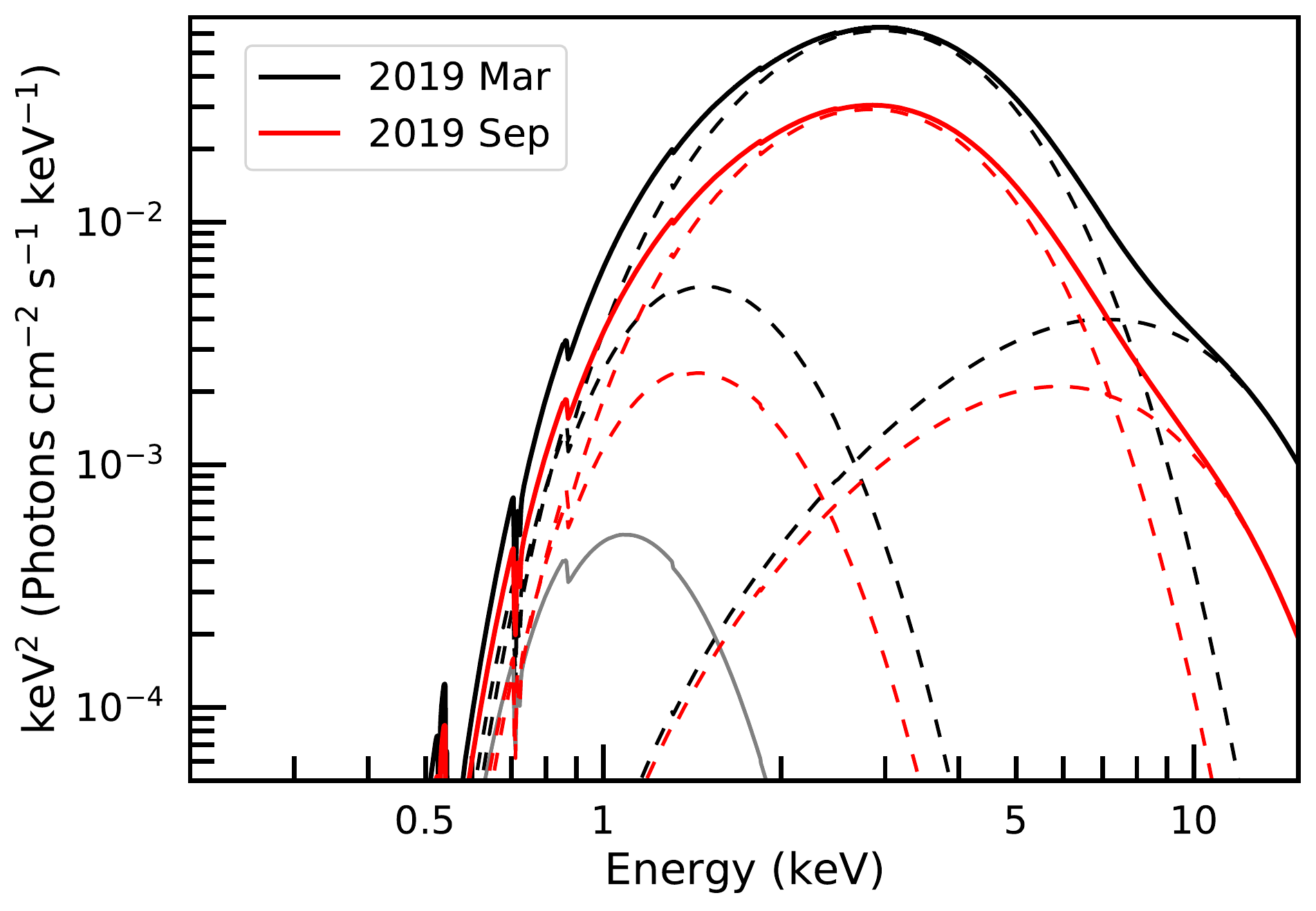}
    \caption{{\it Top}: The broad-band  $E^2 f(E)$ unfolded spectra of \src\ from the simultaneous \xmm\ and \nus\ observations performed in 2019 March (black) and September (red). The best-fitting 4BB models are plotted with a solid line (see Sec.\,\ref{sect:spectral} for more details). Post-fit residuals in units of standard deviations are shown in {\it Panel b}. {\it Panel c}: Post-fit residuals after setting the normalization of the BB component accounting for the whole NS to zero. {\it Panel d}: Post-fit residuals after setting the normalization of the cold BB component to zero. {\it Panel e}: Post-fit residuals after setting the normalization of the warm BB component to zero. {\it Panel e}: Post-fit residuals after setting the normalization of the hot BB component to zero. {\it Bottom}: The $E^2 f(E)$ plot of the fitted models, with the contributions of the single components. In grey, we plot the BB component that accounts for the emission from the whole stellar surface.  
    }
    \label{fig:spectra_4bb}
\end{figure}

Firstly, we fit the \xmm\ and \nus\ spectra, extracted from nearly simultaneous observations, performed on 2019 March 4 and September 22--28. For the first \xmm\ pointing, we limited the fitting energy range to 0.7--9\,keV, where events are optimally calibrated\footnote{\label{fn:note1}\url{https://xmmweb.esac.esa.int/docs/documents/CAL-TN-0018.pdf}} for the EPIC-pn in timing mode. While for the second one, the range was extended to 0.3--10\,keV. Both observations provided high-statistical-quality spectra where systematic calibration uncertainties are important. Therefore, we added an energy
independent systematic uncertainty of 2\% to each spectral
channel\footref{fn:note1}. For \nus\ spectra, we limited the analysis to the 3--15\,keV energy range because of the very low signal-to-noise ratio of \src\ above 15\,keV. Following \citet{2009A&A...498..195B}, we applied a three-blackbody spectral model (3BB; {\sc bbodyrad} in {\sc Xspec}), assuming that one of the thermal components could be identified with the emission from the whole neutron star surface. We fit the spectra jointly, including a renormalization constant to account for cross-calibration uncertainties between the two instruments. All parameters of the 3BB model were left free to vary between the two different epochs except for the hydrogen column density \nh, which was forced to maintain the same value in all spectra. The fit gave an overall satisfactory description with \nh\ = (8.7$\pm$0.3)$\times10^{21}$\,\cmcm\ and a reduced chi-squared \rchisq=1.07 for 613 degrees of freedom (dof). 
The best-fitting values relative to this model are listed in Table \ref{tab:spectra_3bb}. We note that the inferred parameters for the coldest BB component, \rcold $\sim$5.5\,km and \ktcold $\sim$0.24\,keV, are different from the values estimated by \citet{2009A&A...498..195B} for the thermal emission from the whole neutron star surface ($R_{\rm NS}\sim$12.8\,km\footnote{\citet{2009A&A...498..195B} assumed a distance of 3.5\,kpc. We scaled the value for the new distance measurement.} and $kT_{\rm NS}\sim$0.144\,keV). This result hints at the presence of a new cold component during the latest outburst, besides the hot and warm ones. In order to investigate further this hypothesis, we fit the spectra with a 4BB model. The parameters of the additional BB component were frozen at $R_{\rm NS}$=12.8\,km and $kT_{\rm NS}$=0.144\,keV, as derived for the previous outburst \citep{2009A&A...498..195B}. This component accounts for the emission from the entire stellar surface and we are assuming that it is not affected by the outburst, as it occurred during the 2003 event. The fit yields \nh\ = (9.7$\pm$0.2)$\times10^{21}$\,\cmcm\ (\rchisq=1.06 for 613 dof). The best-fitting values are listed in Table \ref{tab:spectra_4bb}, where the \nus\ results refer to the data acquired with the FPMA. We repeated all the analysis using the FPMB spectra and the results were compatible within the uncertainties with those reported in Table \ref{tab:spectra_4bb}. Figure \ref{fig:spectra_4bb}, top panel, shows the 2019 March and September spectra with the best-fitting 4BB model, the residuals with respect to this model and the residuals after setting to zero the normalization of each BB component to highlight the contribution of each spectral component to the source spectrum.

In the following, we study in detail all the \nus\ observations performed during this first outburst stage. The \nus\ spectra start from 3\,keV, we then decided to fix the \nh, $R_{\rm NS}$ and $kT_{\rm NS}$ at the values derived in Table \ref{tab:spectra_4bb}. Moreover, we kept the radius and temperature of the cold BB component frozen at the values derived from the simultaneous, broad-band fit of \xmm\ and \nus\ data (see Table\,\ref{tab:spectra_4bb}). For the three \nus\ spectra relative to pointings carried out before 2019 March, we fixed \rmed\ and \ktmed\ at the values obtained for the 2019 March data set, assuming the cold BB component to be constant through the first $\sim$80 days of the outburst.  

The first \nus\ pointing performed soon after the radio re-activation of the source (ID.90410368002, epoch: 2018 Dec) was already presented by \citet{2019ApJ...874L..25G}. We reanalysed it in a consistent way with our approach. Source emission was detected up to $\sim$30\,keV. The 4BB model revealed structured residuals above $\sim$15\,keV, therefore an extra component was required to describe the detected hard tail. The addition of a power law (PL) to the 4BB model provided a good description of the data with \rchisq=0.9 for 135 dof. Best-fitting parameters were: \ktmed=0.59$\pm$0.02\,keV, \rmed=3.5$\pm$0.6\,km, \kthot=0.96$\pm$0.06\,keV, \rhot=0.66$^{+0.24}_{-0.17}$\,km, and photon index $\Gamma$=2.1$\pm$0.3\footnote{We did not include this first observation in Table\,\ref{tab:spectra_4bb_nustar} because of the additional power law needed to adequately fit the spectrum. We do not find evidence of this non-thermal hard component in the following data sets.}. 
We note that \citet{2019ApJ...874L..25G} fitted a different model to this spectrum over the range 3--30 \,eV (i.e., a BB+2PL model), deriving a hard power-law photon index of $\Gamma_h \sim$ --0.3 for the PL characterizing the non-thermal emission, which becomes dominant above $\sim$20\,keV. 
The four following \nus\ observations were acquired between January and September 2019. The spectral analysis was limited to the 3--15\,keV energy interval because of the very low source signal-to-noise ratio above 15\.keV. The spectra were well-fitted by the 4BB model, with no need of the extra PL component required at the outburst peak. The fading of the non-thermal component during the outburst decay is typically observed in magnetar outbursts \citep[e.g.,][]{2009MNRAS.396.2419R}. The best-fit parameters are reported in Table\,\ref{tab:spectra_4bb_nustar}. During the first year of the outburst decay, we can already notice a softening of the source emission from the \xmm\ and \nus\ data, except for the BB component which accounts for the whole surface emitting region with parameters fixed at the quiescent values ($kT_{\rm NS}$=0.144\,keV and $R_{\rm NS}$=12.8\,km). The \ktcold\ and \ktmed\ do not show significant variability, while \kthot\ cools from $\sim$1.5\,keV to $\sim$1.4\,keV. Finally, we measure a shrinking for all the three regions. 

To study the evolution of the hot BB component, we consider the results derived from the fit of the four \nus\ spectra from 2019 January till September with a 4BB model, applying the procedure explained above. In particular, we note that when fitting together the March 2019 \xmm\ and \nus\ data, the value of the \kthot\, is not compatible with that derived from fitting the \nus\ data sets, nor it is compatible with the cooling of such component. To check further on this issue, we did several tests: a) looked at the spectrum using only PATTERN=0 events, and b) extracted the MOS2 (Timing Mode) spectrum, and fit it simultaneously with the \nus\ spectrum. The PATTERN restriction did not change the fit parameters, however when using the MOS2 spectrum the best-fitting parameters for the hot BB component for the 2019 March epoch were \kthot =1.68$\pm$0.07\,keV and \rhot =0.07$\pm$0.02\,km. The derived temperature is lower that what we found using the EPIC-pn (Timing Mode) data (Table\,\ref{tab:spectra_4bb}). We then ascribe the small difference between the March 2019 value of \kthot\, in Table\,\ref{tab:spectra_4bb} and Table\,\ref{tab:spectra_4bb_nustar} to uncertainties in the cross-calibration between \nus\, and the EPIC-pn in Timing Mode. We used the values derived by fitting the \nus\, long-term monitoring to study the cooling in time of the hot component.

\begin{figure*}
    \centering
    \includegraphics[scale=0.5]{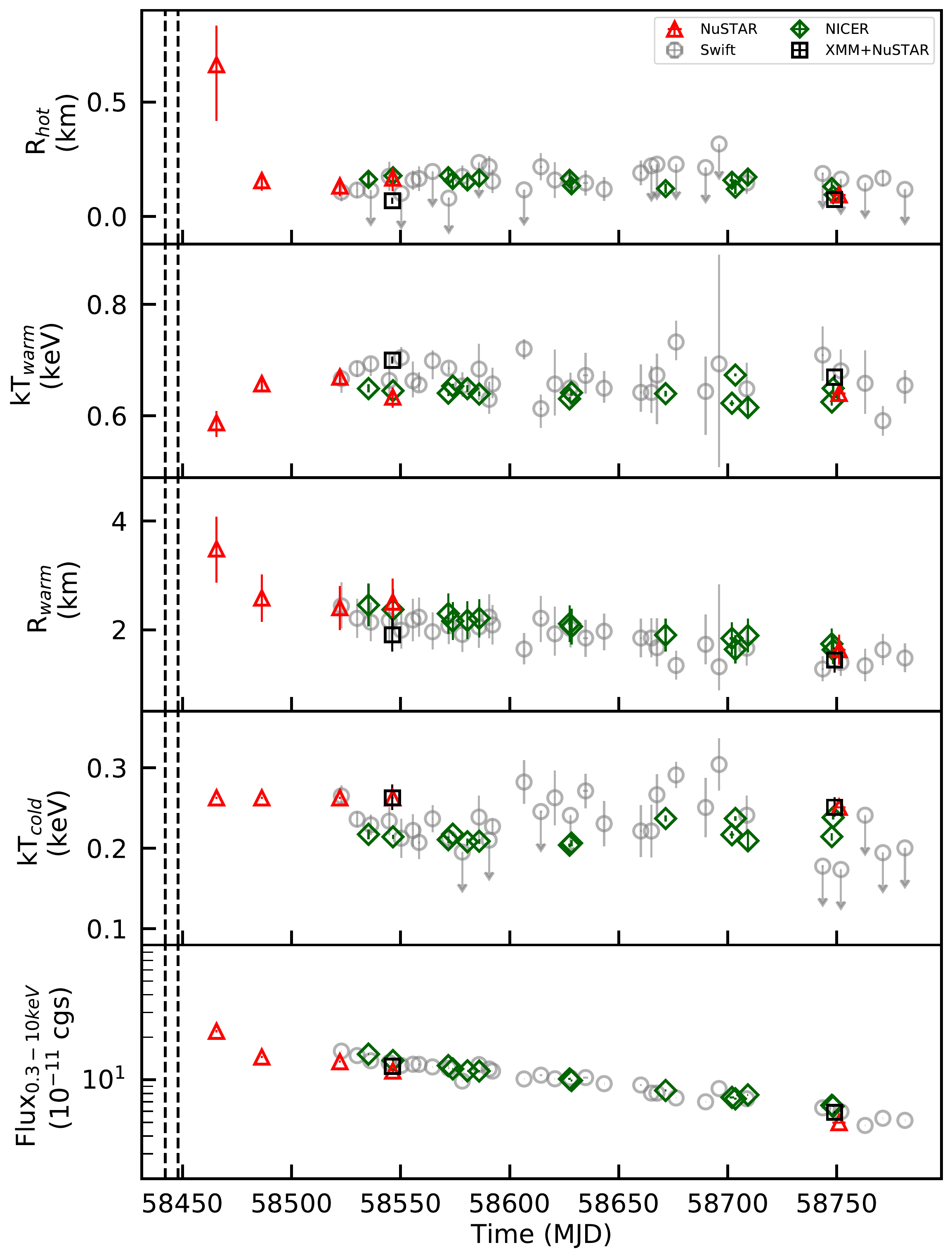}
    \caption{Temporal evolution of the free spectral parameters for the absorbed 4BB model (for details see Section\,\ref{sect:spectral}). BB radii are evaluated for an observer at infinity and a distance of 2.5\,kpc. The bottom panel shows the temporal evolution of the observed flux in the 0.3--10\,keV energy band. The dashed vertical lines denote the range of the outburst onset as constrained by \citet{2019ApJ...874L..25G}, 2018 November 20--26 (MJD 58442--58448).}
    \label{fig:spec_evo}
\end{figure*}

To study the source evolution in time we also added the \swift/XRT and the \nicer\ observations. The \swift/XRT spectra were fit in the 0.8--10\,keV energy band for WT mode observations owing to known calibration issues at lower energies\footnote{\url{http://www.swift.ac.uk/analysis/xrt/digest_cal.php}} and in the 0.3--10\,keV interval for PC mode data sets, whereas in the \nicer\ pointings the source signal-to-noise ratio was higher in the 0.6--7\,keV energy range. To increase the photon statistics, we merged observations carried out a few days apart after verifying that no significant flux variations were present (see Table\,\ref{tab:log}). We hence focused on 26 \swift/XRT WT spectra, 5 \swift/XRT PC spectra and 14 \nicer\ spectra. For consistency, we fit all the data sets with a 4BB model, where \nh\ and \rcold\ were held fixed at the values of the \xmm+\nus\ spectra closer in time, and \kthot\ at the values derived from the \nus\ spectra (Table\,\ref{tab:spectra_4bb_nustar}). The temporal evolution of the spectral parameters is shown in Figure \ref{fig:spec_evo}. The bottom panel of Figure \ref{fig:spec_evo} shows the observed flux corresponding to the 4BB model as a function of time in the 0.3--10\,keV energy range. 
The first observation performed on 2018 December 13 caught the source at a flux of (2.2$\pm$0.3)$\times10^{-10}$\,\flux, a factor of $\approx$415 greater than the quiescent value recorded by \R\ in 1993, (5.3$\pm$0.5)$\times10^{-13}$\,\flux\
(\citet{2018MNRAS.474..961C}; see also Table\,3 by \citet{2004ApJ...605..368G}).
During the following $\sim$320 days, the flux decreased slowly and reached a value of (5.2$\pm$0.2)$\times10^{-11}$\,\flux\ on 2019 October 25, which is still about two orders of magnitude above quiescence.

\begin{figure*}
    \centering
    \includegraphics[width=1.\columnwidth]{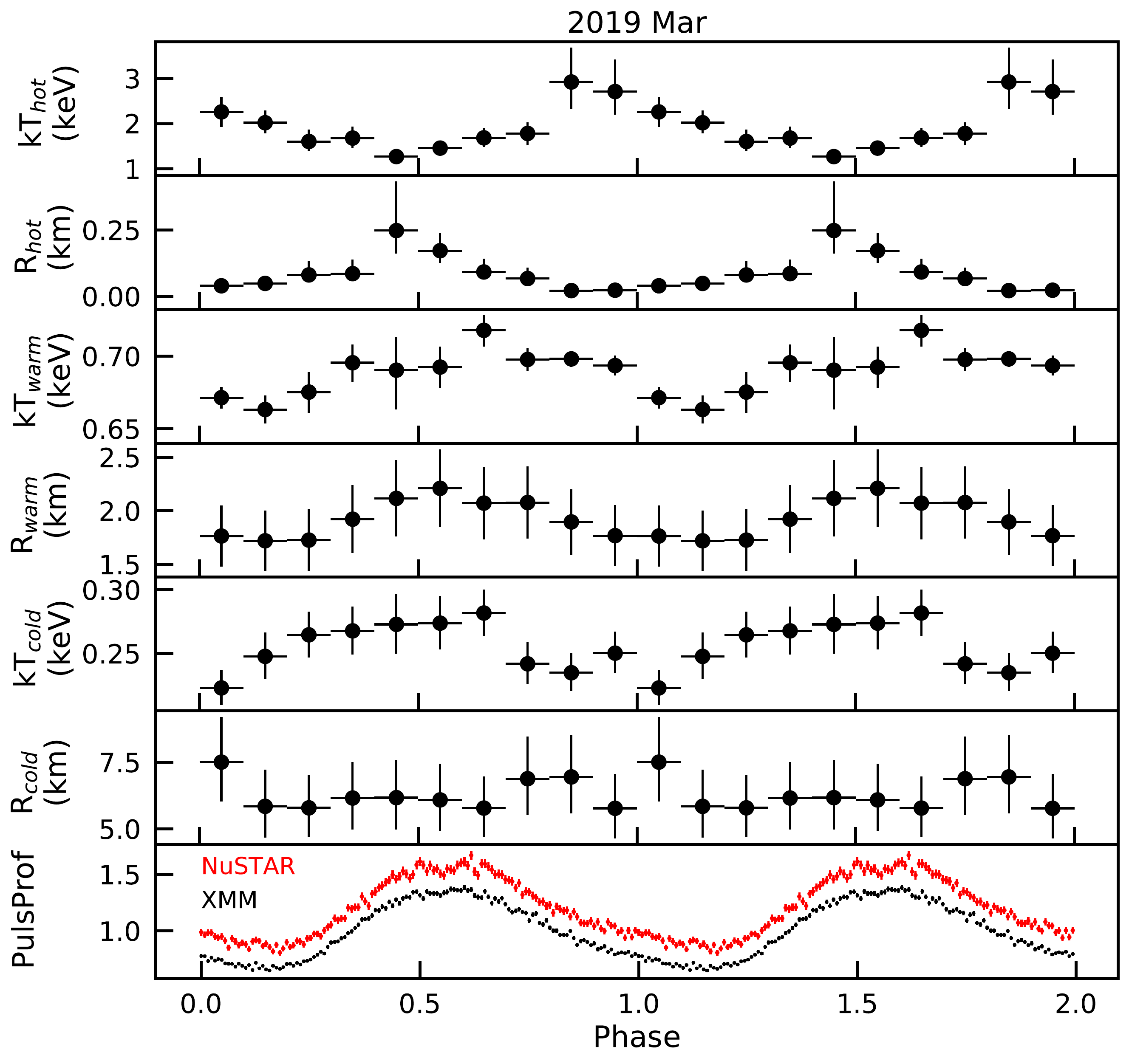}
    \includegraphics[width=1.\columnwidth]{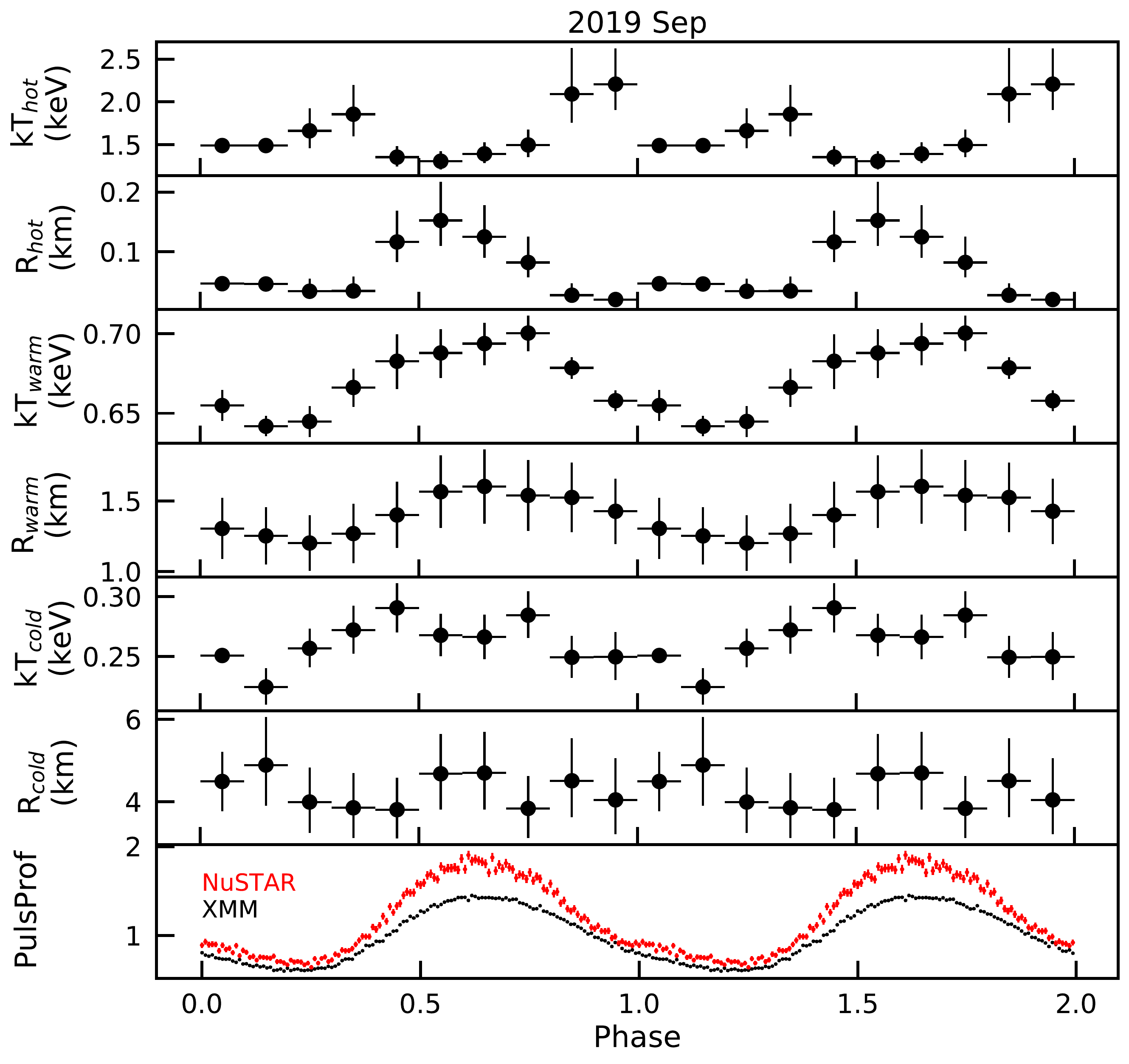}
    \caption{Phase evolution of the cold, warm and hot thermal components for the (quasi-)simultaneous \xmm+\nus\ observations performed in 2019 March (left) and September (right), with the corresponding pulse profiles (bottom panel). For plotting purpose, the \nus\ profiles have been shifted along the vertical axis and two phase cycles are shown.}
    \label{fig:pps}
\end{figure*}

\begin{figure*}
    \centering
    \includegraphics[width=2.\columnwidth, trim= 0 1.cm 0 1.5cm, clip]{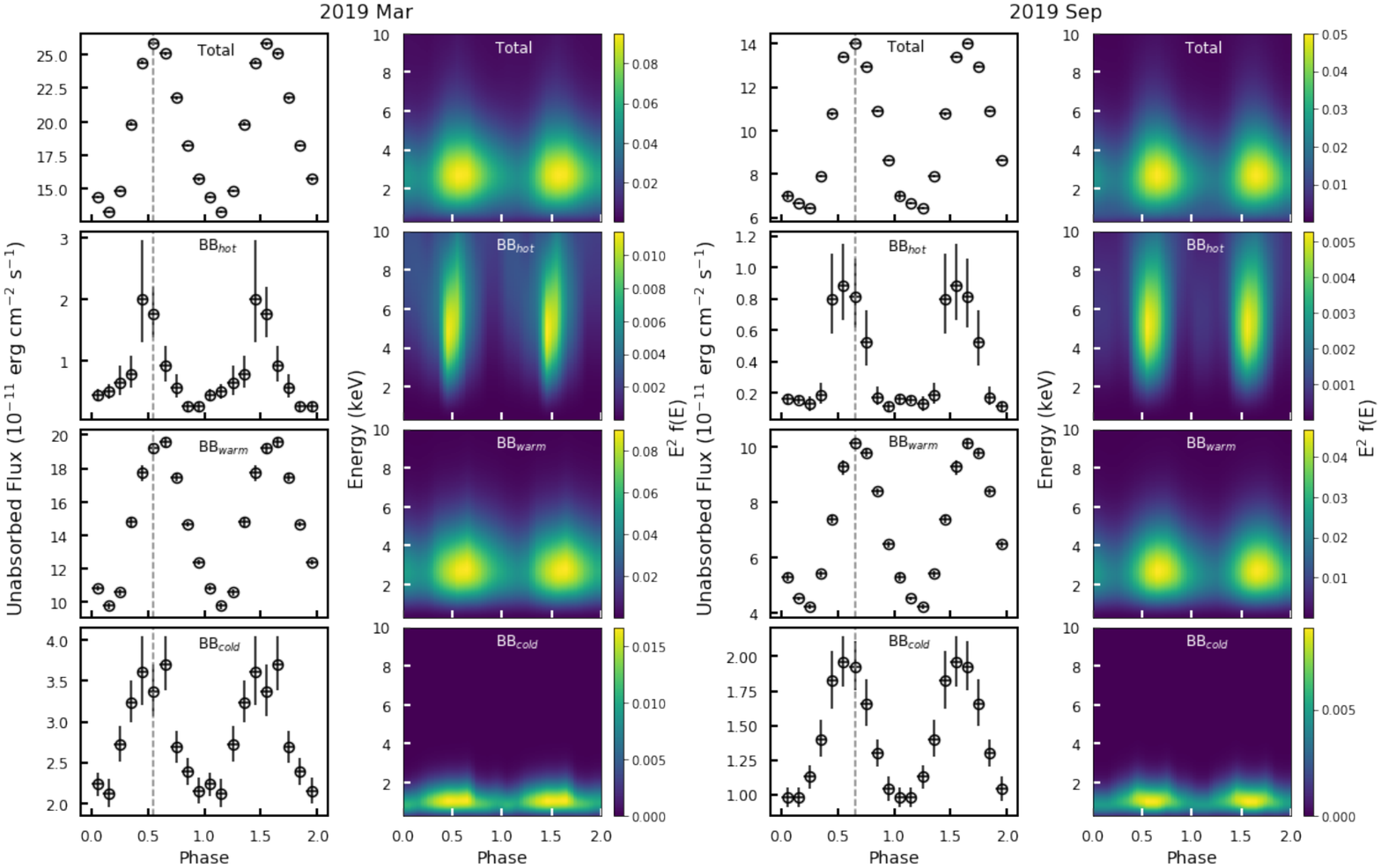}
    \caption{Dynamic spectral profiles for 2019 March (left) and September (right) observations. For each data set, the first column shows the 0.3$-$10\,keV unabsorbed flux for the total model, and hot, warm, and cold blackbody components (from top to bottom). The second column shows the contour plots in the phase-energy plane for the total, hot, warm, and cold blackbody $E^2 f(E)$ flux. The colour scale is in units of keV$^2$\,(photons\,cm$^{-2}$\,s$^{-1}$\,keV$^{-1}$).}
    \label{fig:dps}
\end{figure*}

\subsection{Phase-resolved spectroscopy and dynamic spectral profiles}
\label{sect:phase_resolved}

We then carried out a pulse-phase resolved spectroscopic analysis of the (quasi-)simultaneous \xmm\ and \nus\ observations performed in 2019 March and September. The phase cycle was divided into 10 phase intervals in order to rely on a large enough number of photons. The phase-resolved spectra were fit simultaneously with the 4BB model.
The column density was held fixed at the phase-averaged value (\nh\ = 9.7 $\times$ 10$^{21}$\,\cmcm; see Section\,\ref{sect:spectral}). The radius and the temperature for the BB component accounting for the emission from the whole neutron star were frozen at $R_{\rm NS}$=12.8\,km and $kT_{\rm NS}$=0.144\,keV, while all the other parameters were allowed to vary.   
The fits gave \rchisq=1.04 for 1907 dof for the 2019 March data set and \rchisq=1.01 for 1719 dof for the 2019 September spectra. We found a variability of all the free parameters through the whole cycle, as shown in Figure\,\ref{fig:pps}.

To examine the spectral variation as a function of the star rotational phase, we produced the Dynamic Spectral Profiles (DSPs), shown in Figure\,\ref{fig:dps}, for both epochs.  
The DSPs show the contour plots of the $E^2 f(E)$ flux as a function of phase and energy, derived from the 10 phase-resolved spectra extracted as explained above. The top panel refers to the total flux, obtained from the 4BB model, while the panels below display the flux of the single thermal components ($BB_{\rm hot}$, $BB_{\rm warm}$, $BB_{\rm cold}$). 

From the DSPs and the evolution of the unabsorbed flux as a function of phase (see Figure\,\ref{fig:dps}), it is evident that the $BB_{\rm hot}$ component appears to be shifted in phase with respect to the total and the $BB_{\rm warm}$ emission for both epochs. 

\subsection{Bursting behaviour}
We searched for short bursting activity in the \xmm, \nus\ and \nicer\ observations.
Four short $\sim$1\,s X-ray bursts were previously reported by \citet{2005ApJ...629..985W}, in the months following \src's 2003 outburst.
These bursts were characterized by short spikes of $\sim$1\,s followed by extended tails of 100\,s of seconds on enhanced flux.
We find that there is no evidence of similar short bursts of this type in the data sets analyzed in this paper.
Bursting activity has also been recently reported by \citet{2020arXiv200508410P} searching through \nicer\ data from observations between MJDs 58520 and 58540.
\citet{2020arXiv200508410P} reported on thousands of smaller bursts, with widths less than or equal to one rotational cycle, using a `zero crossing' algorithm, rather than flux enhancement above the mean rate.
We searched for this bursting activity in each \xmm, \nus\ and \nicer\ observation, by calculating the mean pulse profile for each observation,
and computing the differences between each individual X-ray pulse and the mean pulse profile. 
We grouped these count rate differences together according to rotational phase, in order to look for changes in flux from the mean flux at a given rotational phase.
Then, at each rotational phase, we compared the distribution of the differences in counts with the expected distribution from Poisson sampling. We find that the differences in count rates are consistent with Poisson fluctuations.

\section{Simultaneous radio observations}
\label{sect:radio}

We observed \src\ with the Sardinia Radio Telescope (SRT, \citealt{bolli,prandoni}) on 2019 February 8 for 3.6\,hr starting at UT 08:01:40 (Project ID 08-19), simultaneously with \nus\footnote{The \nus\ observation ID.80202013003 started on 2019 February 7 at 15:31:09 TT and ended on 2019 February 8 at 14:56:09 TT and provided a total overlap of $\sim 2$ hr with the SRT observation.}. Data were recorded with the ATNF digital backend Pulsar Digital Filter Bank\footnote{See \url{http://www.srt.inaf.it/media/uploads/astronomers/dfb.pdf}.} (PDFB), in search mode over a usable bandwidth of 924\,MHz centered at 6.5 GHz, with a spectral resolution of 1\,MHz. Full Stokes data were 2-bit sampled every 0.5\,ms.

\begin{figure}
    \centering
    \includegraphics[width=8truecm]{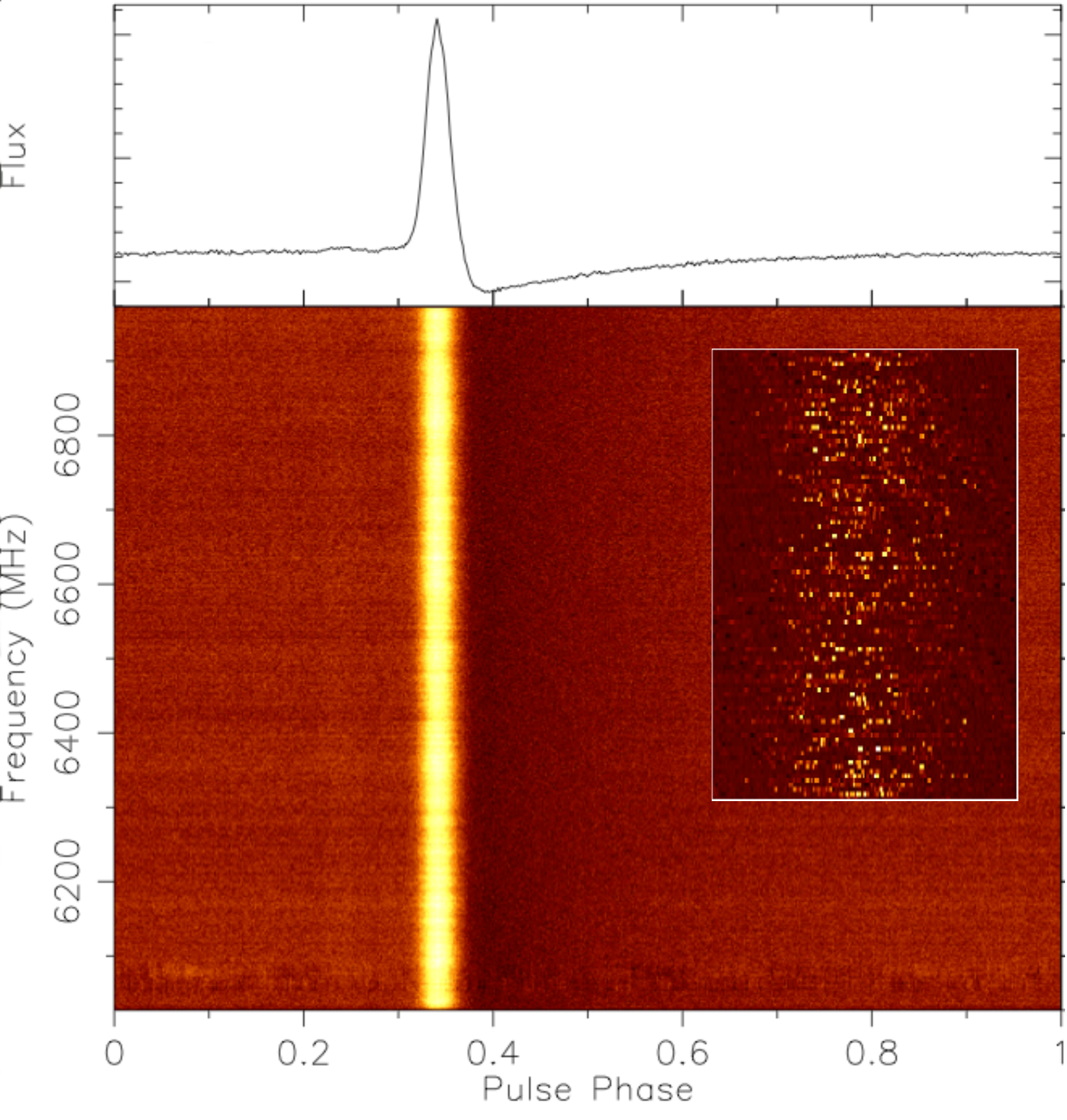}
    \caption{Frequency versus rotational phase (bottom) for the integrated pulse profile (top) of the 3.6 hr SRT observation.  The dispersion delay is removed. The inset shows a subset of 100 single pulses plotted in the phase range 0.31 to 0.37.}
    \label{fig:radio}
\end{figure}

The data were folded over 1024 time bins using the ephemeris obtained from the \nus\ simultaneous observation, showing bright pulsed emission, as it was also reported e.g., by \citet{2019MNRAS.488.5251L} for various frequencies between 1.4 and 8.4 GHz. A waterfall plot of the integrated pulse profile over the entire observing bandwidth is shown in Figure \ref{fig:radio} (bottom), and the single peaked profile, with a $\sim200$ ms width, is plotted at the top. The optimized signal-to-noise ratio of the pulse $\rm S/N = 1066$ implies a flux density of $\sim2.5$\,mJy, assuming an antenna gain of $0.6$\,K/Jy and a system temperature of $\sim$30\,K. The estimated flux density is about three times lower than the value derived by \citet{2019MNRAS.488.5251L} on 2019 January 24. The radio emission was bright enough that the pulses were visible above the noise at every rotation. A snapshot of 100 subsequent single pulses stacked in phase, is plotted in the inset of Figure\,\ref{fig:radio}. The vast majority of single pulses are detected within the phase range of the main peak, but a minority ($<$ 1\%) appears about 0.1 earlier in rotational phase. A very dim precursor of the main peak can indeed be seen in the waterfall plot. Single de-dispersed pulses show multiple narrow components with a high degree of linear polarization, as it had been also seen by e.g., \citet{serylak09} at 1.4, 4.8 and 8.4 GHz. No X-ray bursts were detected simultaneously with any of the radio pulses nor at any time during the 2019 February \nus\ observation. Moreover, we compared the phase alignment of the X-ray and radio pulse profiles. We found that the X-ray peak lags the radio one by $\sim$0.08 cycles, behaviour consistent with that observed at the peak of this outburst \citep{2019ApJ...874L..25G} and during the 2003 outburst \citep{2016ApJ...820..110C}.

\section{Discussion}
\label{sect:disc}

We have presented the evolution of the spectral and timing properties of the magnetar \src\ during its second outburst which started in November 2018, about fifteen years after its discovery outburst in 2003. Our monitoring campaign covered the first $\sim$320 days of the outburst decay, allowing us to characterise accurately the behaviour of the source over a long time-span. In the last observation, performed on 2019 October 25 ($\sim$11 months after the outburst onset), the observed 0.3--10\,keV flux was (5.2$\pm$0.2)$\times$10$^{-11}$\,\flux, still two orders of magnitude higher than the historical quiescent level measured in archival \R\ data taken in 1993 \citep{2018MNRAS.474..961C,2004ApJ...605..368G}.\\ 

\subsection{Light curve modelling}
\begin{figure*}
    \centering
    \resizebox{\hsize}{!}{\includegraphics{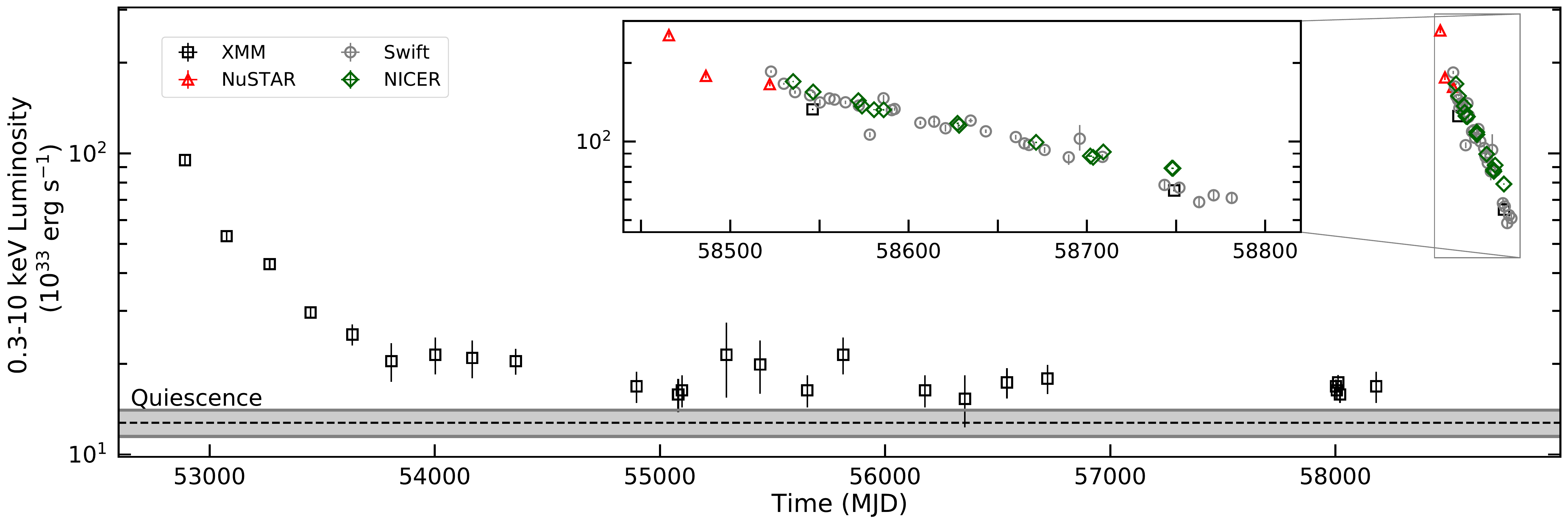}}
    \caption{Long-term evolution of the 0.3--10\,keV luminosity of \src\ from \xmm\ (black squares), \nus\ (red triangles), \swift\ (grey circles) and \nicer\ (green diamonds) data acquired between 2003 September and 2019 October. The solid black line denotes the quiescent level (1.3$\times10^{34}$\,\lum) derived by \citet{2018MNRAS.474..961C} from the \R\ observation performed on 1993 April 3 and scaled for the new distance measurement. The grey area represents the associated uncertainty. The inset is a zoom of the 2018 outburst.}
    \label{fig:long_LC}
\end{figure*}

To study the 0.3--10\,keV luminosity decay, we adopt a simple phenomenological model which consists of a combination of a constant and an exponential function:  
\begin{equation}
    L(t) = L_{\rm q} + (L_{\rm max} - L_{\rm q}) \exp{(-(t-t_0)/ \tau)},
\end{equation}
where $L_{\rm q}$ is the quiescent luminosity, $L_{\rm max}$ is the luminosity at the outburst peak, $t_0$ is the onset time of the outburst and $\tau$, the $e$-folding time, can be considered as an estimate of the decay timescale. We fixed $L_{\rm q}$ to the quiescent level derived by \citet{2018MNRAS.474..961C} from the \R\ observation performed in 1993 and scaled for the new distance measurement, (1.3$\pm$0.1)$\times10^{34}$\,\lum. The onset of the outburst was constrained to be between 2018 November 20--26 \citep[MJD 58442--58448;][]{2019ApJ...874L..25G} and we assumed the mid epoch, 2018 November 23 (MJD 58445), as $t_0$. Therefore, only ($L_{\rm max} - L_{\rm q}$) and $\tau$ were allowed to vary in the fit. A 10 per cent error was assigned to each luminosity. The best-fitting values are ($L_{\rm max} - L_{\rm q}$)=(2.1$\pm$0.1)$\times10^{35}$\,\lum\ and $\tau$=223$\pm$11\,d. We computed the energy released in the outburst by integrating the best-fitting model for the light curve over the whole duration of the event. Our decay fit predicts that the source will reach a luminosity level consistent with the quiescent value around 2022 January, releasing a total energy of $\sim$4$\times10^{42}$\,erg. This value is estimated assuming no change in the decay pattern and, hence, should be considered only as a rough estimate. Figure\,\ref{fig:long_LC} shows the long-term light curve of \src, spanning from 2003 September till 2019 October. The source underwent a previous outburst in 2003, whose onset was missed and is constrained to be between 2002 November 14 and 2003 January 23 \citep[MJD 52595$-$52662;][]{2004ApJ...609L..21I}. A monitoring campaign started on 2003 September, and the first observation caught \src\ at a luminosity level of $\sim 9.5\times$10$^{34}$\,\lum, a factor of $\sim$7 higher than the pre-outburst level. At the 2018 outburst onset, the luminosity reached a value of $\sim$2.5$\times$10$^{35}$\,\lum, slightly higher than what was measured in the previous outburst and a factor of 20 above the historical minimum. After $\sim$320 days, the luminosity decreased till $\sim$6$\times$10$^{34}$\,\lum, a factor $\sim$5 higher than the quiescent level.  

\subsection{Interpretation of the data with a physical model}
During the monitoring campaign, \src\ showed a thermal spectrum modelled by an absorbed four-blackbody model, apart from the first observation performed close to the outburst onset when a power-law tail was detected up to $\sim$30\,keV ($\Gamma$=2.1$\pm$0.3). One of the blackbody component accounts for the emission coming from the whole neutron star and its parameters are fixed at the values inferred from the \R\ spectra when the source was in quiescence, $kT_{\rm NS}$=0.144\,keV and $R_{\rm NS}$=12.8\,km \citep[see e.g.,][and references therein]{2009A&A...498..195B}. Besides this last component, we identified three thermally emitting areas on the star. After an initial decrease, the radius of the hot component \rhot\ settled to a constant value of $\sim$0.1\,km; whereas the temperature \kthot\ cooled from $\sim$1.5\,keV to $\sim$1.4\,keV (see Table\,\ref{tab:spectra_4bb_nustar}). As shown in Figure\,\ref{fig:spec_evo}, the warm region shrunk from \rmed $\sim$2.6\,km till $\sim$1.5\,km during the $\sim$320 days of monitoring, while the corresponding temperature attained a value in the range 0.6$-$0.7\,keV. For the cold area, the temperature \ktcold\ did not show variability and the radius \rcold\ decreased from $\sim$6\,km to $\sim$4.5\,km (see Table\,\ref{tab:spectra_4bb}). 

The simplest scenario one can envisage for the geometry of the emitting region consists of three circular, concentric, zones, superimposed to the colder star surface: a hot cap at \kthot, surrounded by a warm corona at \ktmed, which is in turn surrounded by a colder ring at \ktcold. This is similar to the picture adopted by \cite{2009A&A...498..195B}, although only two thermal components, a hot and a warm one in addition to emission from the entire surface, were needed in the 2003 outburst \citep[see also][]{2010ApJ...722..788A}. At the peak of the 2003 event, the hot and warm regions reached lower temperatures than those measured during the 2018 outburst (\ktmed$_{2003}$ $\sim$ 0.3\,keV and \kthot$_{2003}$ $\sim$ 0.7\,keV). Moreover, the size of the hot/warm areas varied during the 2003 outburst, showing a steady decrease, while the temperatures remained nearly constant. 

Besides these differences, observations suggest that the emission geometry in the 2018 outburst was indeed more complex. 
As discussed in section \ref{sect:timing} (see in particular Figure \ref{fig:pfshift}), in fact, the pulsed fraction (PF) increases with increasing energy and, even more relevant, the phase of the maximum of the pulse profile also changes with energy. The magnitude of this phase variation is typically $\sim 5$--$10\%$ and depends on the epoch, but the general trend always exhibits an increase up to $\approx 3$\,keV, followed by a decrease. Three concentric caps can not produce such a behaviour since the phase of the maximum is always the same at different energy bands, irrespective of the viewing geometry.

\begin{figure}
    \centering
    \includegraphics[width=11truecm]{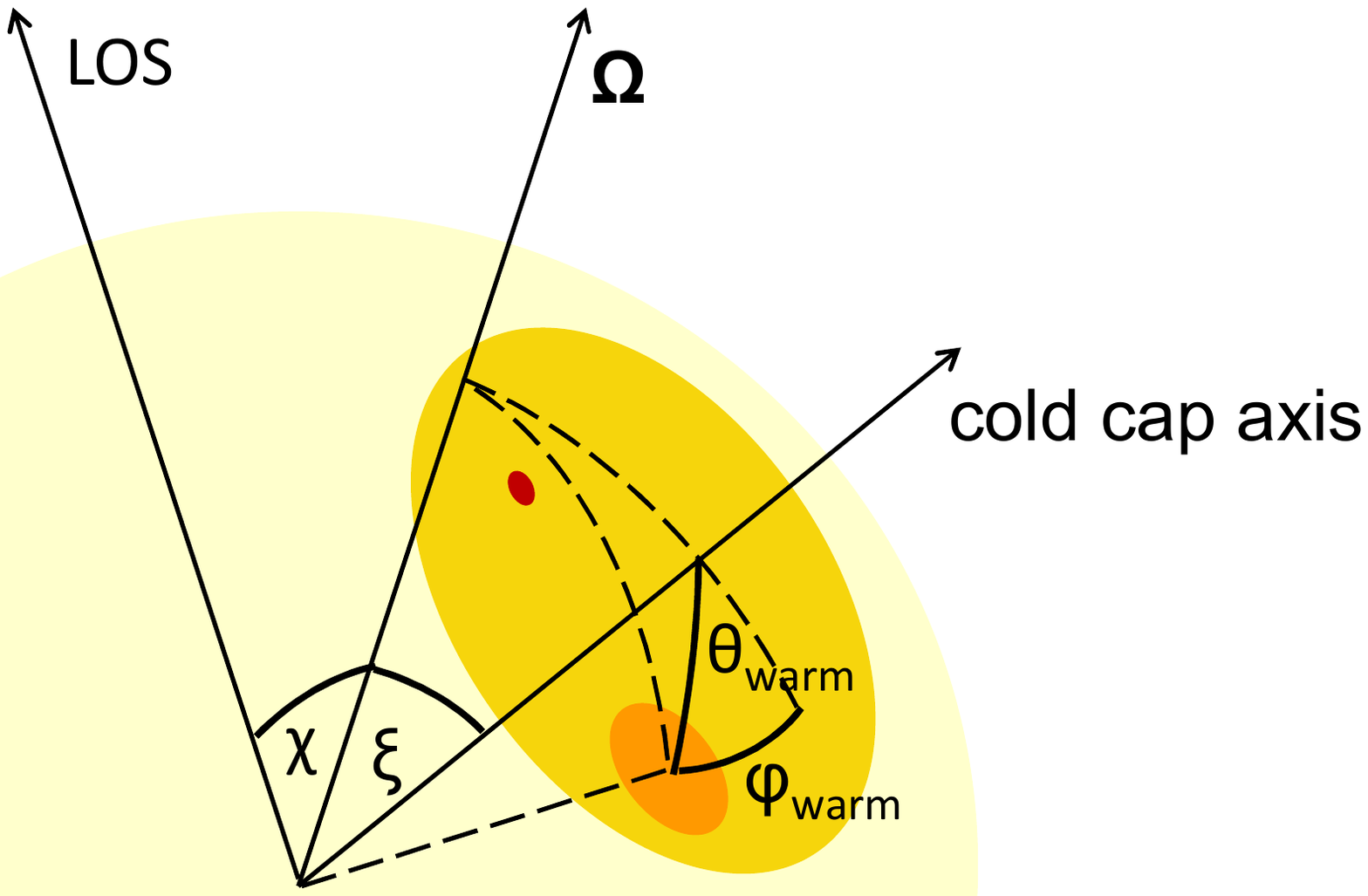}
    \caption{Schematic representation of the surface map of \src\ depicting the different thermal emitting regions: the non-concentric hot (red) and warm (orange) caps, inside the cool one (yellow), superimposed to the colder star surface. The two angles fixing the warm cap position, $\theta_\mathrm{warm}$ and $\phi_\mathrm{warm}$, are shown, together with the two geometrical angles $\xi$ and $\chi$; here $\mathbf{\Omega}$ is the spin axis.}
    \label{fig:map}
\end{figure}

In order to check if a simple, geometrical model can reproduce the observations, we consider again blackbody emission from three regions, but allowing the relative positions of the three caps to vary. The caps are taken to be circular and at constant temperature. Furthermore, we restrict ourselves to the case in which the two hotter caps are inside the cooler one (other cases can be also dealt with). The rest of the surface is at temperature $kT_{\mathrm NS}=0.14$ keV. The model parameters are the cap angular semi-apertures, temperatures (these are fixed by the observed blackbody values for a star radius of $13$ km) and the relative positions with respect to the axis passing through the centres of the cold cap and of the star; these are expressed through the two pairs of angles $\theta_\mathrm{hot,warm}$ and $\phi_\mathrm {hot,warm}$. In addition, there are two angles which fix the overall geometry, $\chi$ and $\xi$, between the line-of-sight and the spin axis and the cold cap axis and the rotation axis, respectively (see Figure\,\ref{fig:map}).

\begin{figure}
    \centering
    \includegraphics[width=8truecm]{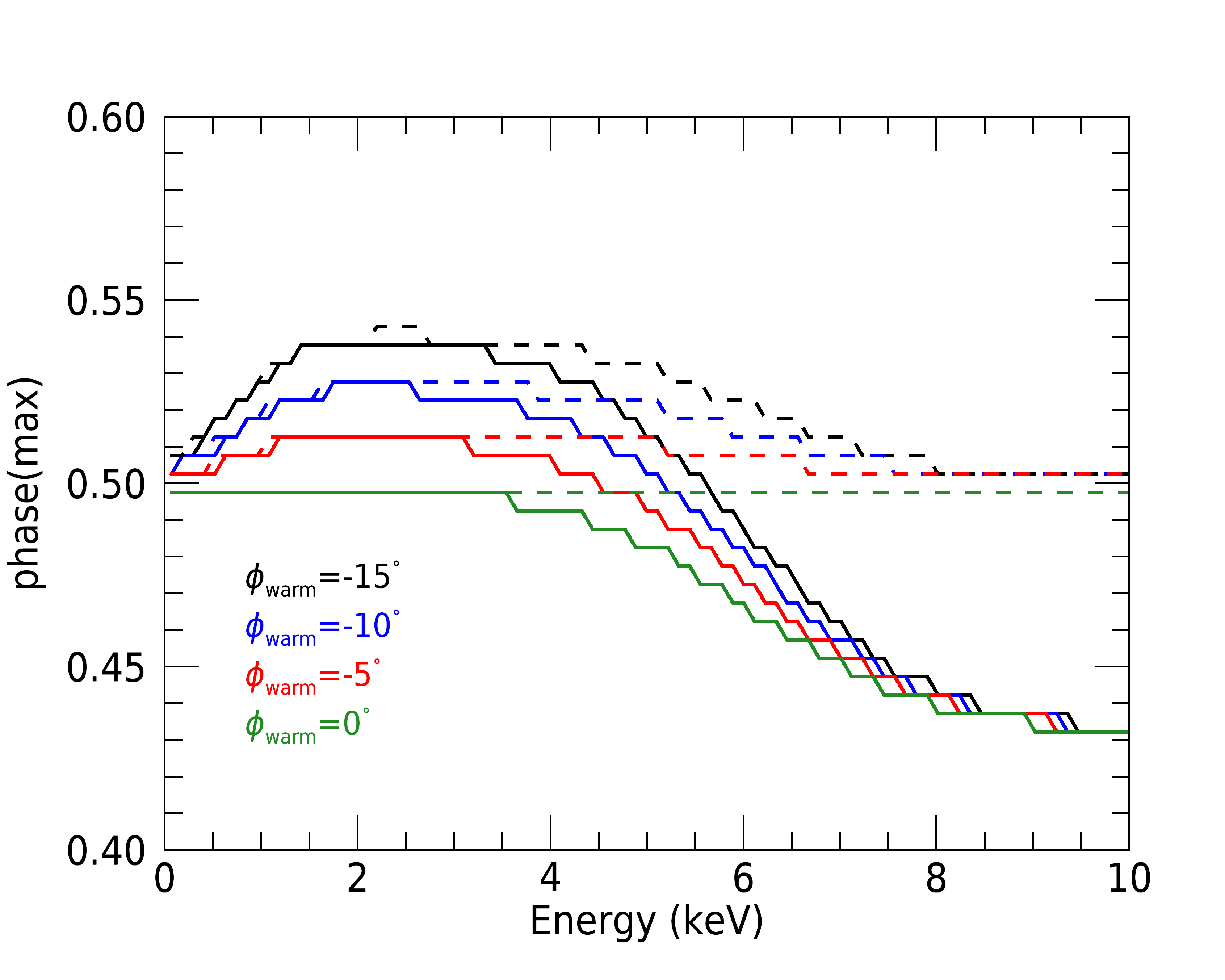}
    \caption{Phase of the maximum of the pulse profile vs. energy (at infinity) for different values of the azimuthal displacement of the hot and warm caps. The full lines are for $\phi_\mathrm {hot}=25^\circ$ and the dashed ones for $\phi_\mathrm {hot}=0^\circ$. See text for details.}
    \label{fig:maxphase}
\end{figure}

We computed phase-resolved spectra using the method described in \cite{2013ApJ...768..147T}, which includes general-relativistic ray-bending; the star mass and radius are $M_\mathrm{NS}=1.4\,M_\odot$ and $R_\mathrm{NS}=13$ km.  
The variation of the phase of the pulse profile maximum with the energy is shown in Figure~\ref{fig:maxphase} for
a given viewing geometry ($\xi =10^\circ$, $\chi =80^\circ$), fixed meridional cap shifts ($\theta_\mathrm{hot}=5^\circ$, $\theta_\mathrm{warm}=-5^\circ$) and different values of $\phi_\mathrm{hot}$ and $\phi_\mathrm{warm}$.

The observed trend and amplitude of variation is indeed recovered, provided that the hot cap precedes the warm one ($\phi_\mathrm{hot}>0$ and $\phi_\mathrm{warm}<0$); if the signs are switched the curve has a minimum in place of a maximum. This behaviour is quite independent on both the assumed values of the meridional shifts and of the angles $\xi$ and $\chi$ and convincingly shows that the thermal structure of the heated region on the star surface is asymmetric. Actually, the  modulation of the blackbody temperatures with rotational phase (see section \ref{sect:phase_resolved} and Figure \ref{fig:pps}) is indicative of more complex thermal map, possibly a single heated region with a (continuous) temperature variation across it, of which our three-temperature model is just an approximation. Indeed, the appearance of such (asymmetric) thermal configurations in response to impulsive energy deposition in the star crust was been recently predicted using 3D simulations \citep[][]{2020ApJ...903...40D}.

\begin{figure}
    \centering
    \includegraphics[width=1.2\columnwidth, trim = 4cm 0 0 0, clip ]{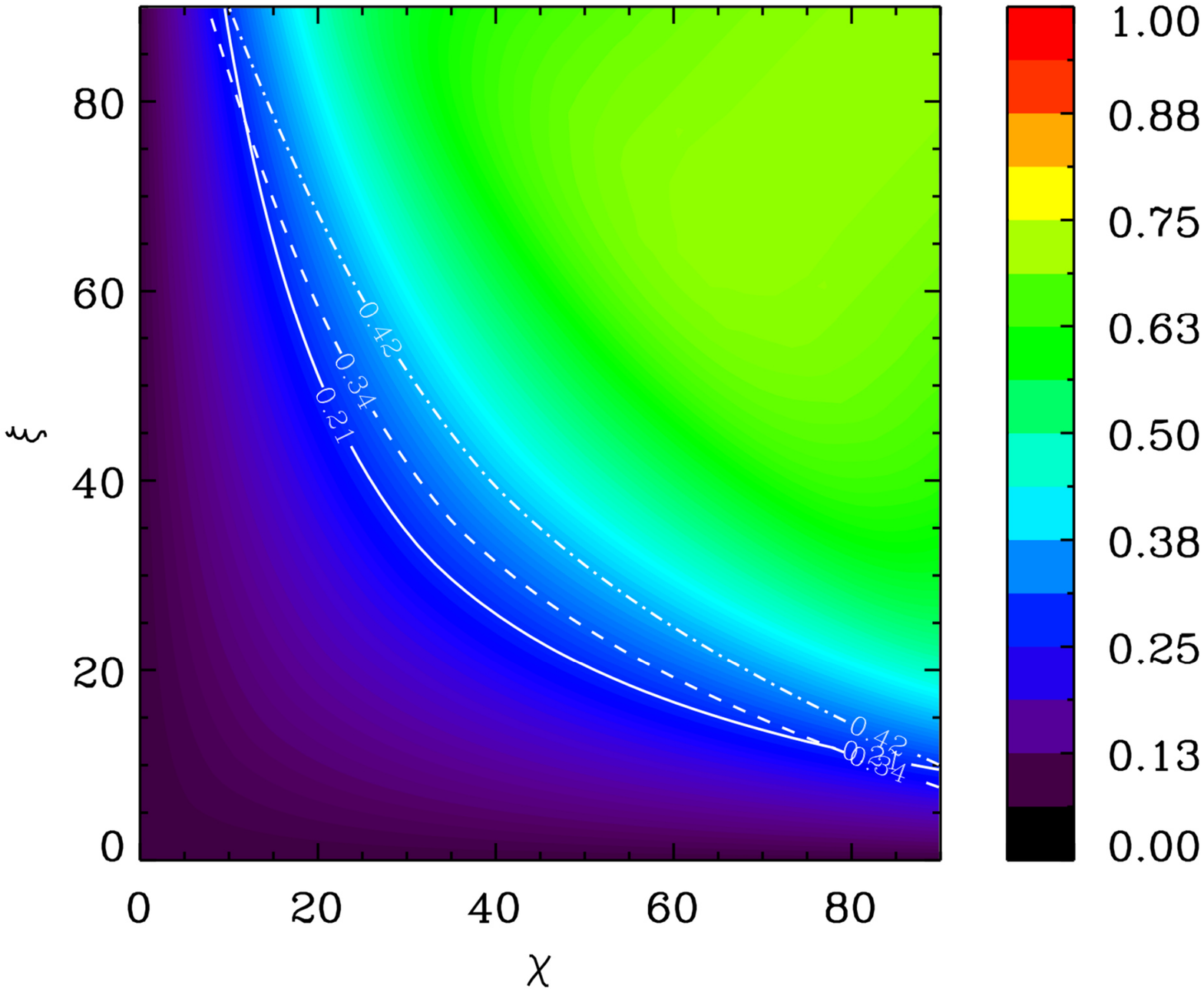}
    \hspace{28pt}
    \caption{Pulsed fraction as a function of the two geometrical angles $\xi$, $\chi$ for a (red-shifted) energy of $0.6$ keV. The white lines mark the loci where the predicted PF matches the measured values for the 2019 March observation at $0.6$ (full), $2.5$ (dashed) and $4.2$ keV (dash-dotted). See text for details.}
    \label{fig:pfchixi}
\end{figure}

The geometry of the thermal emission of \src\ during the 2003 outburst has been studied by various authors \citep{2008ApJ...681..522P, 2010ApJ...722..788A,  2011MNRAS.418..638B}. 
These works derived similar values of $\xi\sim 23^\circ$, while $\chi$ was either $\sim 53^\circ$ or $\sim 148^\circ$\footnote{\cite{2010ApJ...722..788A} considered also resonant Compton scattering in the magnetosphere, so that their model is not purely thermal like that of \cite{2008ApJ...681..522P}.}. We computed pulse profiles for our model at different energies, assuming the same parameter values discussed above, with $\phi_\mathrm{hot}=23^\circ$ and $\phi_\mathrm{warm}=-15^\circ$, but for all possible combinations of the two angles $\xi$ and $\chi$. The pulse profiles are quasi-sinusoidal and single-peaked. Results for the PF are shown in Figure \ref{fig:pfchixi} where the constant PF contours are plotted for a (red-shifted) energy of $0.6$ keV. The white lines mark the loci where the predicted PF matches the measured values for the 2019 March observation at $0.6$ (full), $2.5$ (dashed) and $4.2$ keV (dash-dotted). The PF indeed increases with energy, as expected since harder photons come from a smaller area. Clearly, the observed values of the PF in the different bands should be reproduced for the same ($\xi$, $\chi$) pair for the model to work, i.e. there should be at least one point in the plot of Figure \ref{fig:pfchixi}  where the three lines intersect. This condition is not exactly met in the present case, although there is an indication that small $\xi$ ($\sim 10^\circ$) and large $\chi$ ($\sim 80^\circ$), or the opposite, can provide the correct answer. In this respect we note that our analysis is meant to provide an idea rather than a detailed fit to observations and, as such, contains a number of approximations. Besides the inherent simplification of assuming a three temperature cap instead of a realistic temperature map, we did not take into account the detector response function nor atmospheric effects or radiative bending in computing the pulse profiles. Moreover, all measurement errors were neglected, nor we attempted a complete exploration of the model parameter space. Finally, we remark that, if our estimate of the geometrical  angles is viable, the region on the star surface of \src\ involved in the 2018 outburst was likely different from that of the 2003 event.

\subsection{Timing anomalies and spin-down torque variations}
Regarding the timing properties, the pulse profile attained a single-peak shape as observed in quiescence and in the 2003 outburst \citep[see e.g.,][and references therein]{2016MNRAS.458.2088P, 2019MNRAS.483.3832P}. During the first $\sim$320 days of the outburst decay, the 0.3--10\,keV pulsed fraction increased from $\sim$27\% to $\sim$54\%. This trend is opposite to that measured during the previous outburst, when the pulsed fraction decreased from $\sim$50\% till $\sim$20\% over 4 years, but not unusual within the magnetar population (e.g., the pulsed fraction of SGR\,0418+5729 increased for $\sim$900 days since the outburst onset; \citealp{2013ApJ...770...65R}). The different evolution of the pulsed fraction with time also points to a different affected region with respect to the previous outburst (as discussed above).

Given the timing solution derived by \citet{2019MNRAS.483.3832P} during the quiescent phase prior to the 2018 outburst, we can extrapolate the spin period of \src\ at the epoch of the first \nus\ observation. We found that the predicted period is larger than the measured value ($\Delta P$ $\sim$2$\times$10$^{-5}$\,s). This discrepancy might have been caused by the occurrence of a glitch at the time of the (unobserved) outburst onset. \citet{2019ApJ...874L..25G} calculated a glitch magnitude of $\Delta \nu$/$\nu$ = (4.52$\pm$0.15)$\times$10$^{-6}$, a value typical of magnetars (\citet{2014ApJ...784...37D}; see also, e.g., Swift\,J1818.0$-$1607 \citep{2020ApJ...902....1H}). Moreover, we found evidence of spin-down variations during our monitoring campaign: the period derivative measured during the first two months of the outburst decay ($\dot{P} \sim$7.2$\times$10$^{-12}$\,\ss) is a factor of $\sim$2.5 higher than the value in quiescence ($\dot{P} \sim$2.8$\times$10$^{-12}$\,\ss) and a factor of $\sim$2 smaller than the value derived for the rest of its first year evolution ($\dot{P} \sim$1.5$\times$10$^{-11}$\,\ss). This behaviour has already been witnessed in \src\ after the 2003 outburst and appears to be a common feature following magnetar outbursts \citep[e.g., 1E\,1048.1--5937;][]{2020ApJ...889..160A}. Spin-down rate variations associated with outbursts are commonly explained in terms of the untwisting bundle scenario, according to which outbursts are most likely driven by magnetic stresses resulting in twistings of the external magnetic field with the formation of current-carrying localized bundles \citep{2009ApJ...703.1044B}. During the early stage of the outburst, the twist grows, leading to an increase in the number of (open) magnetic field lines crossing the light cylinder and, thus, to a stronger spin-down torque acting on the star. If the twist amplitude is large ($\gtrsim$1\,rad), this effect should start immediately after the twist is implanted. While, if the initial twist amplitude is $<$1\,rad, the spin-down torque may not be affected until the amplitude reaches 1\,rad, which takes time. In this context, the delayed increase in the spin-down torque of \src\ during its recent outburst (with the maximum occurring about 1.5 month after the outburst onset) would be consistent with the progressive growth of a relatively small magnetospheric twist along the early outburst phases. {\rm The presence of a limited twist is also supported by the very short appearance of a spectral power-law tail, which was observed only in the early phase of the outburst.}  At later stages, when the magnetosphere untwists, the spin-down torque is expected to decrease back to the value measured during the quiescent state.    

\section{Conclusions}
\label{sect:conc}

We studied the long-term evolution of the X-ray emission properties of the magnetar \src\ during the first year of its second outburst since its discovery in 2003. During the monitoring campaign, the source showed a thermal spectrum, apart from the outburst peak when a non-thermal tail was detected up to $\sim$30\,keV. A softening of the source emission is evident during the first $\sim$320 days of the outburst decay: the observed 0.3--10\,keV flux decreased from $\sim$2$\times$10$^{-10}$\,\flux\ to $\sim$5$\times$10$^{-11}$\,\flux, about two orders of magnitude higher than the historical minimum. We also reported on the timing properties of \src. The pulse profile always showed a sinusoidal shape and the pulsed fraction increased by a factor of $\sim$2 over $\sim$300 days. At a given epoch, we found that the pulsed fraction increases with energy and the phase of the maximum of the pulse profile also changes with energy. Moreover, we measured spin-down torque variations along the outburst decay. This behaviour and the temperature distribution inferred from the spectral analysis support the scenario in which the outburst is related to the formation and gradual dissipation of a localised twisted bundle of magnetic field lines, analogous to the Solar coronal loops. Finally, we studied the geometry of the emission regions. Observational results hint to the presence of three areas, superimposed to the colder star surface. We considered a scenario according to which there are a hot and a warm cap, which are non-concentric, surrounded by a cooler one. The slippage of the phase with energy is recovered if the hot cap precedes the warm one. In this framework, the observed increase of the pulsed fraction with energy suggests that the source is a nearly aligned rotator seen almost equator on.

\section*{Acknowledgements}
AB thanks Marco Antonelli for useful discussions during the writing of the manuscript. The scientific results reported in this article are based on data obtained with \swift, \xmm, \nicer\ and \nus\ missions, as well as the Sardinia Radio Telescope (SRT). \xmm\ is an ESA science mission with instruments and contributions directly funded by ESA Member States and the National Aeronautics and Space Administration (NASA). \nicer\ is a 0.2--12\,keV X-ray telescope operating on the International Space Station, funded by NASA. \nus\ is a project led by the California Insitute of Technology, managed by the Jet Propulsion Laboratory, and funded by NASA. We made use of data supplied by the UK \swift\ Science Data Centre at the University of Leicester and of the XRT Data Analysis Software developed under the responsibility of the Italian Space Agency (ASI) Science Data Center. The SRT is funded by the Department of University and Research (MIUR), ASI, and the  Autonomous Region of Sardinia (RAS) and is operated as National Facility by the National Institute for Astrophysics (INAF).

AB and FCZ are supported by Juan de la Cierva Fellowship. AB, NR, FCZ, DV, AI and RS are supported by the ERC Consolidator Grant ``MAGNESIA'' (nr. 817661), and acknowledge funding from grants SGR2017-1383 and PGC2018-095512-BI00. RT, MR and SM acknowledge financial support from the Italian MUR through grant ``UNIAM'' (PRIN 2017LJ39LM). EVG and JAJA acknowledge support for this project from NASA grants 80NSSC20K0717, 80NSSC19K1462, 80NSSC20K0046, and 80NSSC21K0129. AP and MB gratefully acknowledge financial support by the research grant {\it iPeska} (P.I. Andrea Possenti) funded under the INAF national call Prin-SKA/CTA approved with the Presidential Decree 70/2016.


\section*{Data availability}

The data underlying this article are public.




\bibliographystyle{mnras}
\bibliography{reference} 



\bsp	
\label{lastpage}
\end{document}